\documentclass[fleqn,10pt]{wlscirep}
\newcommand{\ba}{\begin{eqnarray}}
\newcommand{\ea}{\end{eqnarray}}
\newcommand{\ban}{\begin{eqnarray*}}
\newcommand{\ean}{\end{eqnarray*}}
\newcommand{\braket}[2]{\mbox{$ \langle #1 | #2 \rangle $}}

\newcommand{\ket}[1]{\mbox{$ | #1 \rangle $}}
\newcommand{\bra}[1]{\mbox{$ \langle #1 | $}}

\newcommand{\one}{\leavevmode\hbox{\small1\normalsize\kern-.33em1}}

\begin{document}
\title{Extracting entangled qubits from Majorana fermions in quantum dot chains through the measurement of parity}

\author[1]{Li Dai}
\author[1,2]{Watson Kuo}
\author[1,3,*]{Ming-Chiang Chung}
\affil[1]{Department of Physics, National Chung Hsing University, Taichung, 40227, Taiwan}
\affil[2]{Center of Nanoscience and Nanotechnology, and Institute of Nanoscience, National Chung Hsing University, Taichung 40227, Taiwan}
\affil[3]{Physics Division, National Center for Theoretical Sciences, Hsinchu, 30013, Taiwan}
\affil[*]{Correspondence to M.-C.C (mingchiangha@phys.nchu.edu.tw)}


\begin{abstract}
We propose a scheme for extracting entangled charge qubits from quantum-dot chains that support zero-energy edge modes. The edge mode is composed of Majorana fermions localized at the ends of each chain. The qubit, logically encoded in double quantum dots, can be manipulated through tunneling and pairing interactions between them. The detailed form of the entangled state depends on both the parity measurement (an even or odd number) of the boundary-site electrons in each chain and the teleportation between the chains. The parity measurement is realized through the dispersive coupling of coherent-state microwave photons to the boundary sites, while the teleportation is performed via Bell measurements. Our scheme illustrates \emph{localizable entanglement} in a fermionic system, which serves feasibly as a quantum repeater under realistic experimental conditions, as it allows for finite temperature effect and is robust against disorders, decoherence and quasi-particle poisoning.
\end{abstract}

\flushbottom
\maketitle
%
%

Majorana fermions (MFs), first considered by Ettore Majorana in 1937 for decomposing Dirac fermions into a superposition of real fermions\cite{Majorana-1937}, are hypothetical particles which are their own antiparticles. In particle physics, no elementary particles are MFs except the neutrino whose nature is not explicitly resolved\cite{Majorana-Neutrino}. In condensed matter physics, however, MFs have been proposed as quasi-particle excitations of the $\nu=5/2$ fractional quantum Hall state\cite{fractional-quantum-hall}, at the surface of a topological insulator coupled with a s-wave superconductor\cite{MF-TI}, in the quantum wells or quantum wires\cite{SOI-1,SOI-2,MF-quantum-wire,MF-quantum-wire-2}, and in cold atoms\cite{MF-cold-atom}. Generally, three elements are needed for the realization of MFs\cite{Review-1,Review-2}: strong spin-orbit interaction to generate position or momentum dependent spin directions, superconductivity to induce electrons pairing effect, and Zeeman magnetic field to lift the spin degeneracy. Several experiments have been performed which can be interpreted as emergence of MFs\cite{MF-exp-1,MF-exp-2,MF-exp-3,MF-exp-4}.

MFs are interesting not only due to their fundamental properties but also in the aspect of their potential applications such as topological quantum computation\cite{MF-QC,MF-QC-2}, quantum state transfer\cite{MF-QI-1}, quantum memory\cite{MF-QI-2} and fault-tolerant quantum random-number generation\cite{MF-QI-3}. The unique feature of these applications is the topological phase of matter for which the manipulations in the degenerate ground state subspace are protected against local perturbations that respect the characteristic symmetries of the system e.g. the particle-hole symmetry, and thermal excitations are suppressed by a sizable energy gap\cite{MF-classify,MF-protection,topological-phase}. However, the perturbation that does not respect the symmetries of the system (e.g. unpaired electrons in superconductors) may induce undesirable transitions within the ground state subspace. This is the phenomenon of quasi-particle poisoning\cite{quasi-particle-poisoning-0,quasi-particle-poisoning} which can cause bit-flip errors and decoherence in quantum computation. Therefore, it is important to devise a scheme of quantum computation that not only benefits from the topological properties of the system but also shows robustness against quasi-particle poisoning. This is the motivation of our work.

In this work we propose a scheme for extracting entangled qubits from Majorana fermions through the measurement of parity. The scheme utilizes the topological properties of the system and is also robust against quasi-particle poisoning. The system we consider is two parallel chains of quantum dots as shown in Fig.~1. Each chain is divided into two subchains. We shall demonstrate that, under realistic experimental conditions, each sub-chain encompasses a zero-energy edge mode composed of two unpaired Majorana fermions. The edge mode corresponds to two degenerate ground states, each of which has a definite parity (an even or odd number of electrons). When considering their structure, we find that the inner sites of the sub-chain (the bulk) are generally entangled with the boundary sites (the edges). Moreover, the states for the bulk part in this bulk-edge entangled state also have a definite parity, while for the edges they are themselves (maximally) entangled states between the boundary sites (the parity is definite as well). We propose employing coherent microwave photons to interact dispersively with the edges. In this way, the parity of the edges is measured so that maximally entangled edge states can be extracted. The extraction is robust against quasi-particle poisoning\cite{quasi-particle-poisoning-0,quasi-particle-poisoning}, as will be shown later in the subsection ``Measurement scheme". It can be seen that our proposal illustrates \emph{localizable entanglement}\cite{LE-2} in a fermionic system. Originally,  localizable entanglement is concerned with spin systems and is defined as the maximum amount of entanglement that can be created, on average, between two spins in a spin chain by performing local measurements on other spins. It provides a method to transfer the many-body entanglement to two localized spins. In our work, the spins are replaced by the fermionic sites of quantum dots, and the measurement on other spins is replaced by the coherent parity measurement on the target fermionic sites (the two boundary sites). However, the edge states are not good entangled qubits, because the basis for one subsystem of the states is encoded in a single fermionic site, so that a superposition between the basis states is difficult\cite{qubit-parafermion}. We propose a scheme to transform the edge states into two useful entangled qubits encoded in the boundary sites of the parallel chains so that the superposition of the basis states of the qubit is allowed. The scheme involves a swap operation between the sites of different edge states, and teleportation through the edge states. Fig.~2 and~3 show the flow diagram and principal pulse sequence of our scheme. The fidelity for the entangled qubits can be as high as $0.9$ in the presence of the decoherence induced mainly by the noise of the electrical gate bias\cite{DQD-QC,decoherence-values}. Another common source of decoherence is the environmental charges trapped in the insulating substrate or at the interface of the heterostructure\cite{dephasing-vs-trap}. These random charges interact with the electrons in the quantum dot, which causes severe decoherence. New growth methods for materials with low trapped charge density\cite{low-trapped-charge}, as well as the charge echo techniques\cite{charge-echo}, can alleviate the decoherence.

\begin{figure}
\begin{center}
\includegraphics[width=0.5\textwidth,height=0.2\textwidth]{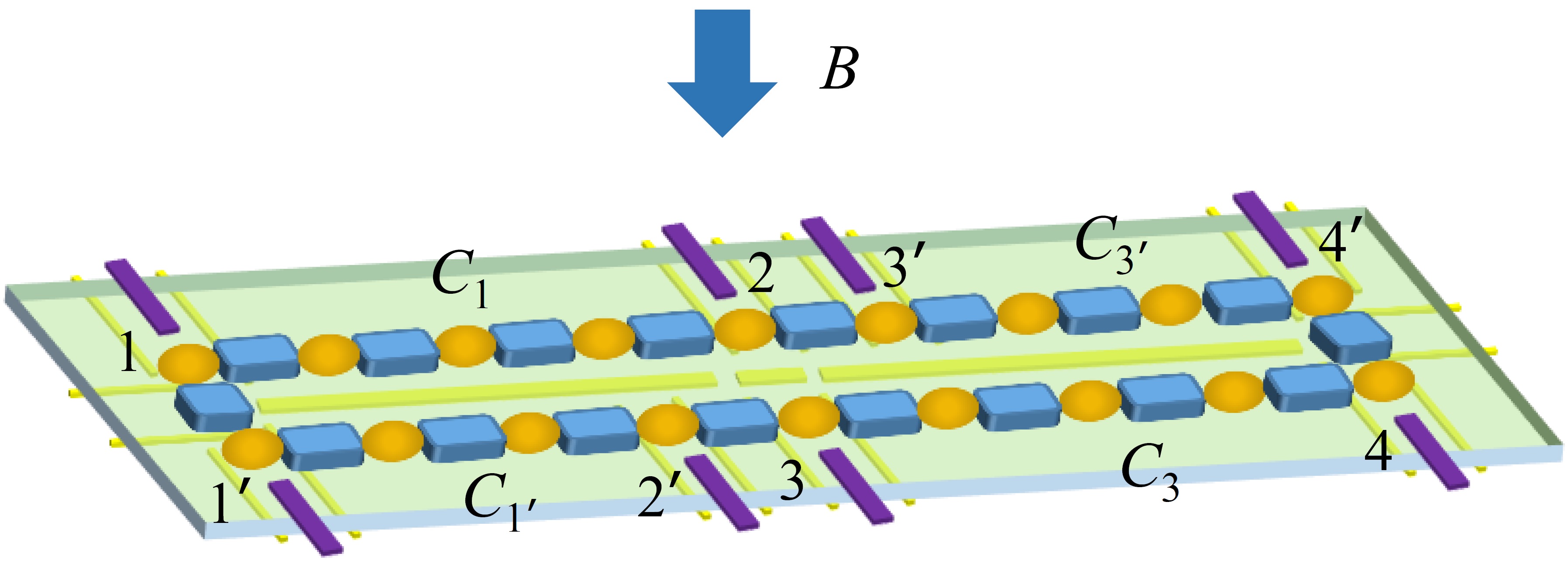}
\end{center}
\caption{\textbf{The schematic diagram of the experimental setup for extracting entangled qubits from two parallel chains that support Majorana fermions. }The quantum dots (dark yellow ellipsoids) and the superconducting grains (blue cuboids) are grown on a dielectric slab (silicon nitride). In the back of the slab, there are electrical gates (golden bars) for controlling the confining potential of the quantum dots as well as the couplings between the quantum dots and the superconducting grains. Each of the two parallel chains are labeled as two connected chains. The chemical potential of their boundary sites are controlled by the front gates (purple). A magnetic field $\textbf{\emph{B}}$ is applied perpendicularly to the chains to induce spin-split levels in the quantum dots.} \label{Fig-1-set-up}
\end{figure}

Our scheme serves as a quantum repeater when combined with the purification protocols which further increase the entanglement\cite{quantum-repeater-purify}. The extracted entangled qubits are a useful entanglement resource in the teleportation-based quantum computation (TQC) which is equivalent to the one-way quantum computation\cite{one-way-QC,Teleport-QC}. TQC can be used as a supplement to the standard charged-based quantum computing (CQC)\cite{DQD-QC} in the situations where quantum gates between remote qubits are needed, with the other ingredients: two- and three-qubit measurements realized by CQC. Our proposal allows for finite temperature effect, as the ground state of the chain is protected by a substantial energy gap induced by the superconducting proximity effect. Also, it is only required to finely tune the system parameters close to the edges, while small disorders of the system in the bulk of the chain is allowed. Moreover, our scheme is robust against quasi-particle poisoning as mentioned in the previous paragraph. Therefore, our proposal can be implemented under realistic experimental conditions.

\begin{figure}
\begin{center}
\includegraphics[width=0.5\textwidth,height=0.25\textwidth]{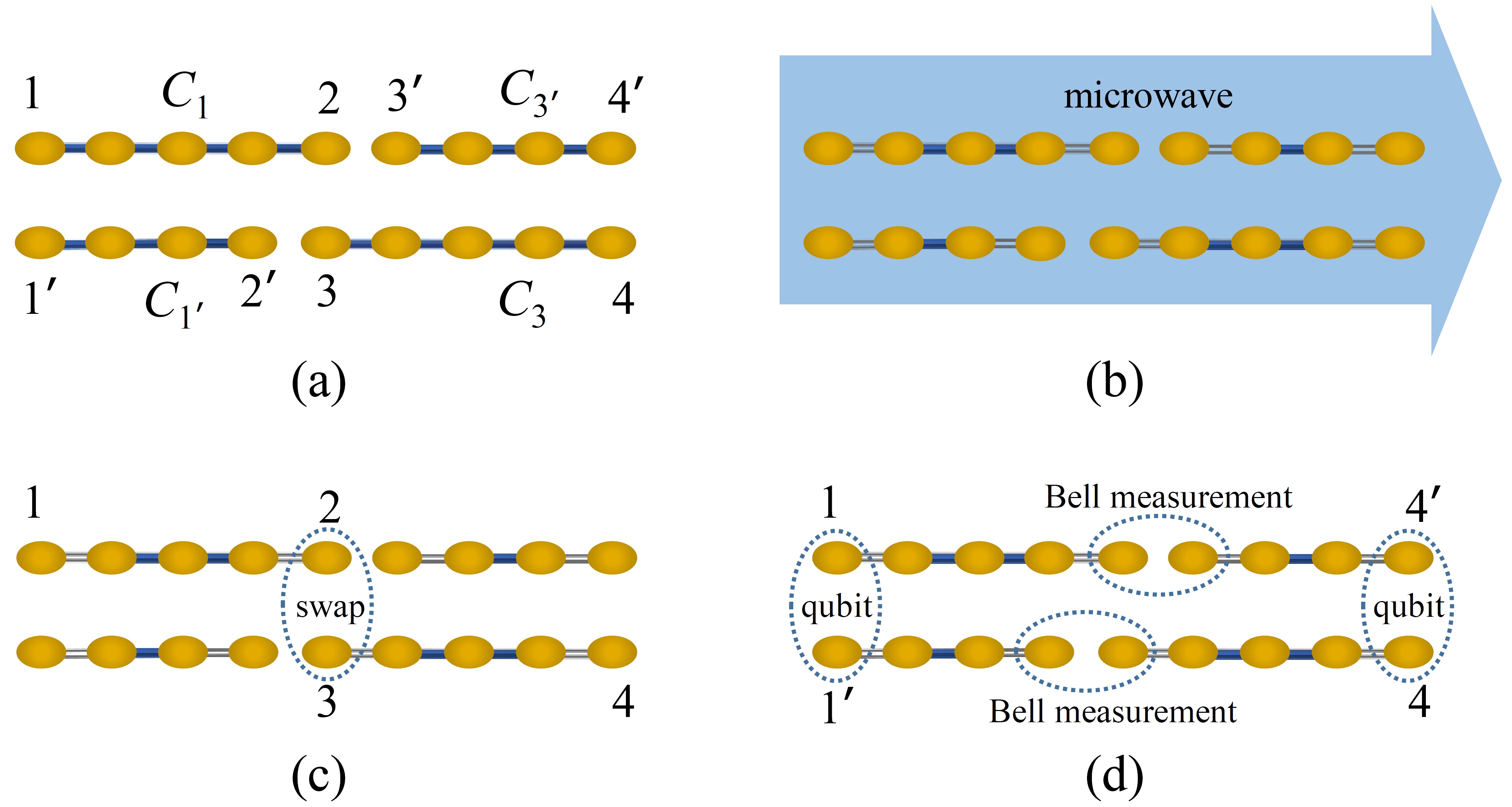}
\end{center}
\caption{\textbf{The flow diagram $(a)\rightarrow (d)$ for extracting entangled qubits from two parallel chains that support zero-energy edge modes. }(a) The two parallel chains, divided into four sub-chains, are initialized in the respective ground states. (b) The microwave is employed to measure the parity of the edges of each sub-chain, in order to collapse the wave function into an entangle state of the boundary sites (see Fig.~5 for details). Here the hollow grey lines indicate that the boundary sites are decoupled from the inner sites. (c) A swap operation between the site $2$ and $3$ is performed, resulting in a bipartitie entangled state where the two subsystems are $(1,3)$ and $(2,4)$. 
(d) Two Bell measurements are performed in order to teleport the state of $3$ to $1'$ and that of $2$ to $4'$. In this way, entangled qubits are formed in the overall ends of the two parallel chains.} \label{Fig-2-flow-diagram}
\end{figure}

\begin{figure}
\centering
\includegraphics[width=0.5\textwidth,height=0.37\textwidth]{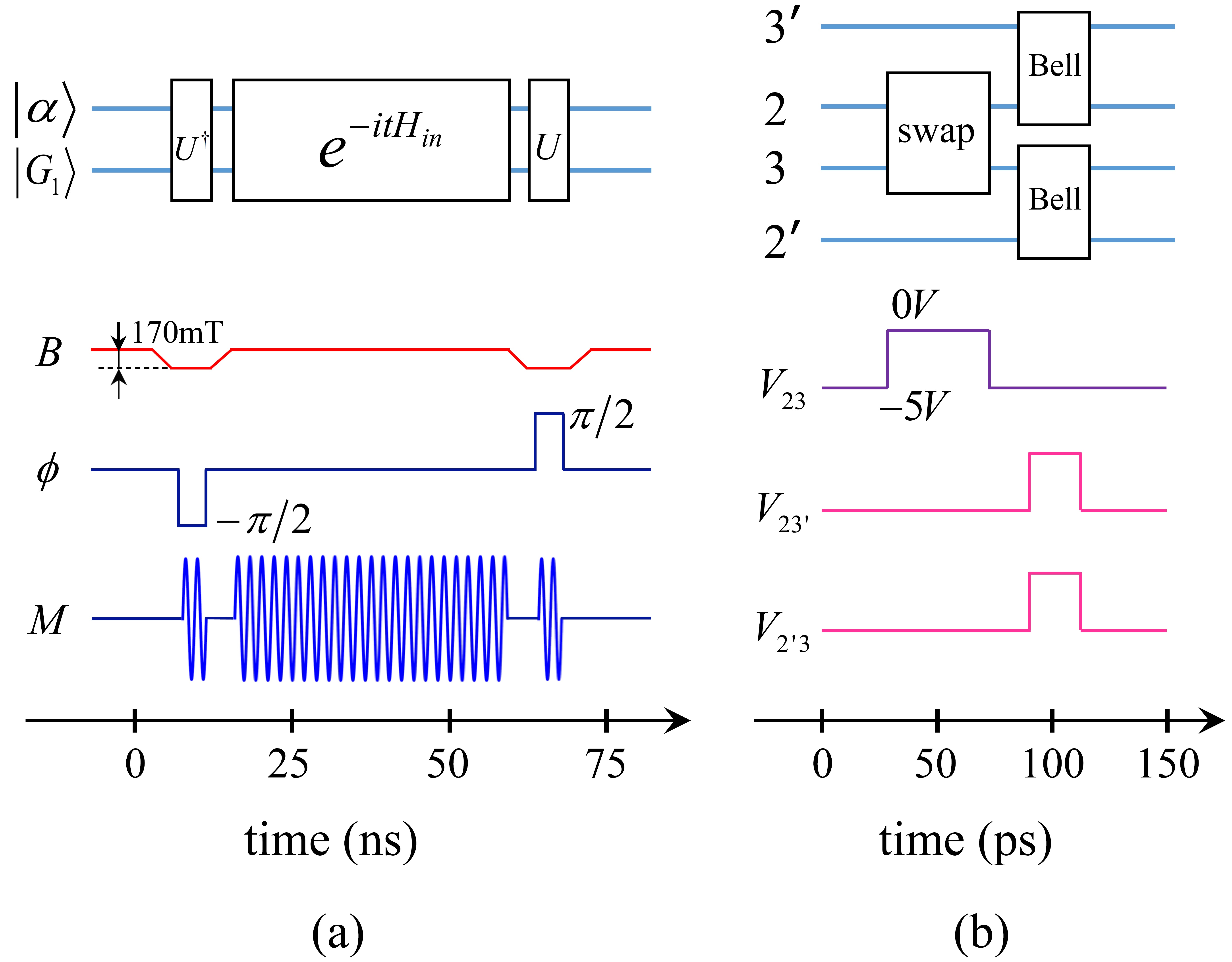}
\caption{\textbf{The schematic diagram of the principal pulse sequence of our scheme for illustrative purposes only. }(a) shows the process of measuring the parity of the boundary-site electrons of one sub-chain. It corresponds to Fig.~2 (b). Here the microwave initially in the coherent state $\ket{\alpha}$ and the sub-chain $C_{1}$ will go through three interactions $U^{\dagger},e^{-itH_{in}},U$ designated by Eq.~(\ref{U-transform}). $B$ is the Zeeman field. Decreasing it by $170$ mT is to tune the gap of boundary-site electrons into resonance with the microwave for realizing $U^{\dagger}, U$. $\phi$ is the phase of the electron-microwave interaction in Eq.~(\ref{microwave-dot-H}) and $M$ is the waveform of the microwave. (b) corresponds to Fig.~2 (c) and (d). Here $V_{jk}$ is the gate voltage between the site $j$ and $k$. Increasing it will decrease the potential barrier, which induces an interaction between the relevant sites.} \label{Fig-3-pulse-sequence}
\end{figure}

\section*{Results}
\subsection*{The model.}For convenience of discussion, each chain in Fig.~1 is re-labelled as two connected sub-chains so that there are four sub-chains $C_{1},C_{1'},C_{3},C_{3'}$, with respective boundary sites being $(1,2), (1',2'),(3,4),(3',4')$. The partition scheme is not unique, as long as the sites $2$ and $3$ are aligned to allow a controllable coupling between them. The system of a single chain has already been proposed by Sau and Sarma for realizing MFs\cite{chain-experiment}.  We briefly review the experimental realization of the chain. The linear chain with $N$ sites is composed of $N$ quantum dots. The two adjacent quantum dots are coupled through a s-wave superconducting grain which induces pairing interaction in the quantum dots\cite{two-dot-chain}. A Zeeman magnetic field $\textbf{\emph{B}}$ is perpendicularly applied to the chain, lifting the spin degeneracy, so that only a single quantum level effectively participates in the interaction between neighboring quantum dots. The chemical potential of each quantum dot is tuned by applying gate voltages individually, in order to resonantly couple the lower spin-split level of quantum dots to the Fermi level of superconductors. The lack of inversion symmetry in quantum dots induces Rashba spin-orbit interaction in the quantum dots, which results in spin texture in the quantum level indispensable for generating proximity effect in the neighboring sites\cite{SOI-1,SOI-2}. The chain is shown\cite{chain-experiment} to support a zero-energy Majorana edge mode for a wide range of system parameters with disorders. The edge mode is topologically protected against local perturbations on the bulk of the chain and thermal noise is also suppressed due to the substantial energy gap of the system\cite{MF-protection,topological-phase}.

We study one sub-chain first. This corresponds to switching off the couplings between any two of the four sub-chains through performing appropriate electrical gating operations. The effective Hamiltonian is\cite{chain-experiment,Kitaev-Majorana}
\ba\label{Kitaev-chain}
H=\sum_{j=1}^{N-1}(-w_{j}c_{j}^{\dagger}c_{j+1}+\Delta_{j}c_{j}c_{j+1}+h.c.)-\sum_{j=1}^{N}\mu_{j}c_{j}^{\dagger}c_{j},
\ea
where the operator $c_{j}^{\dagger}$ ($c_{j}$) creates (annihilates) an electron in the Fermi level (the lower spin-split level with a chemical potential $\mu_{j}$) of the $j$th quantum dot, $w_{j}$ and $\Delta_{j}$ are the tunneling and pairing amplitudes between the $j$th and $(j+1)$th quantum dots, and $h.c.$ denotes the Hermitian conjugation of its previous two terms. The subscripts here only describe the quantum dots in a sub-chain. Their meanings are different from the labelling for the boundary sites in Fig.~1. The parameters $w_{j}$, $\Delta_{j}$ and $\mu_{j}$ can be different from site to site, due to the limited precision in fabricating and controlling quantum dots. We assume $w_{j}>0$, as the phases of $w_{j}$ can be eliminated by an appropriate transformation $c_{k}\rightarrow c_{k}e^{i\theta_{k}}$. Also, $\Delta_{j}$ is chosen to be a real number with a fixed sign (all positive or all negative) through carefully tuning the phases of the superconducting grains\cite{chain-experiment}.

The Majorana operators are defined as follows
\ba\label{majorana-def-1}
d_{2j-1}&=&c_{j}+c_{j}^{\dagger}\,,\nonumber\\
d_{2j}&=&-i(c_{j}-c_{j}^{\dagger})\,,\hspace{0.5cm}j=1,2,\cdots,N.
\ea
It can be seen that they resemble Pauli spin operators but anticommute for different fermionic sites ($d_{j}^{\dagger}=d_{j}, d_{j}^{2}=1, d_{j}d_{k}+d_{k}d_{j}=0$ for $j\neq k$). Eq.~(\ref{Kitaev-chain}) can be written in terms of the Majorana operators,
\ba\label{m-chain}
H=\frac{i}{2}\sum_{j=1}^{N-1}\left[(w_{j}+\Delta_{j})d_{2j}d_{2j+1}+(-w_{j}+\Delta_{j})d_{2j-1}d_{2j+2}\right]-\frac{i}{2}\sum_{j=1}^{N}\mu_{j} d_{2j-1}d_{2j}\,.
\ea
We notice that when
\ba\label{interesting-solution}
\mu_{1}=\mu_{N}=0,\,\,\Delta_{1}=w_{1},\,\,\Delta_{N-1}=w_{N-1},
\ea
the Majorana operators $d_{1}$ and $d_{2N}$ will be absent from Eq.~(\ref{m-chain}).
The two operators form a zero-energy edge mode with its annihilation operator
\ba\label{zero-mode-fast-decay}
\tilde{b}_{N}=\frac{1}{2}(d_{1}+i\eta d_{2N})=\frac{1}{2}(c_{1}+c_{1}^{\dagger}+\eta c_{N}-\eta c_{N}^{\dagger}),
\ea
where $\eta=\det W_{0}$ with $W_{0}$ being a $(2N-2)\times(2N-2)$ real orthogonal matrix that block diagonalizes the coefficient matrix in Eq.~(\ref{m-chain}) (its dimension is reduced by $2$ due to the absence of $d_{1}, d_{2N}$), and $\eta^{2}=1$. The remaining Majorana operators $d_{j},(2\leq j\leq 2N-1)$ form Dirac modes $\tilde{b}_{k},(1\leq k\leq N-1)$ with generally non-zero energies. See Fig.~4. The introduction of $\eta$ in Eq.~(\ref{zero-mode-fast-decay}) is to ensure that the parity operators of the respective Dirac modes are equal: $\prod_{j=1}^{N}(1-2c_{j}^{\dagger}c_{j})=\prod_{j=1}^{N}(1-2\tilde{b}_{j}^{\dagger}\tilde{b}_{j})$. Here the parity operator has two eigenvalues $\pm 1$. The eigenvalue $1$ means the number of electrons (or quasi-particles for $\tilde{b}_{j}^{\dagger}\tilde{b}_{j}$) is even, while $-1$ means the corresponding number is odd. The Dirac modes $\tilde{b}_{k}$ diagonalize Eq.~(\ref{m-chain}): $H=\sum_{k=1}^{N-1}\lambda_{k}(\tilde{b}_{k}^{\dagger}\tilde{b}_{k}-\frac{1}{2})$. See Supplementary Materials for a detailed discussion.

\begin{figure}
\centering
\includegraphics[width=3in]{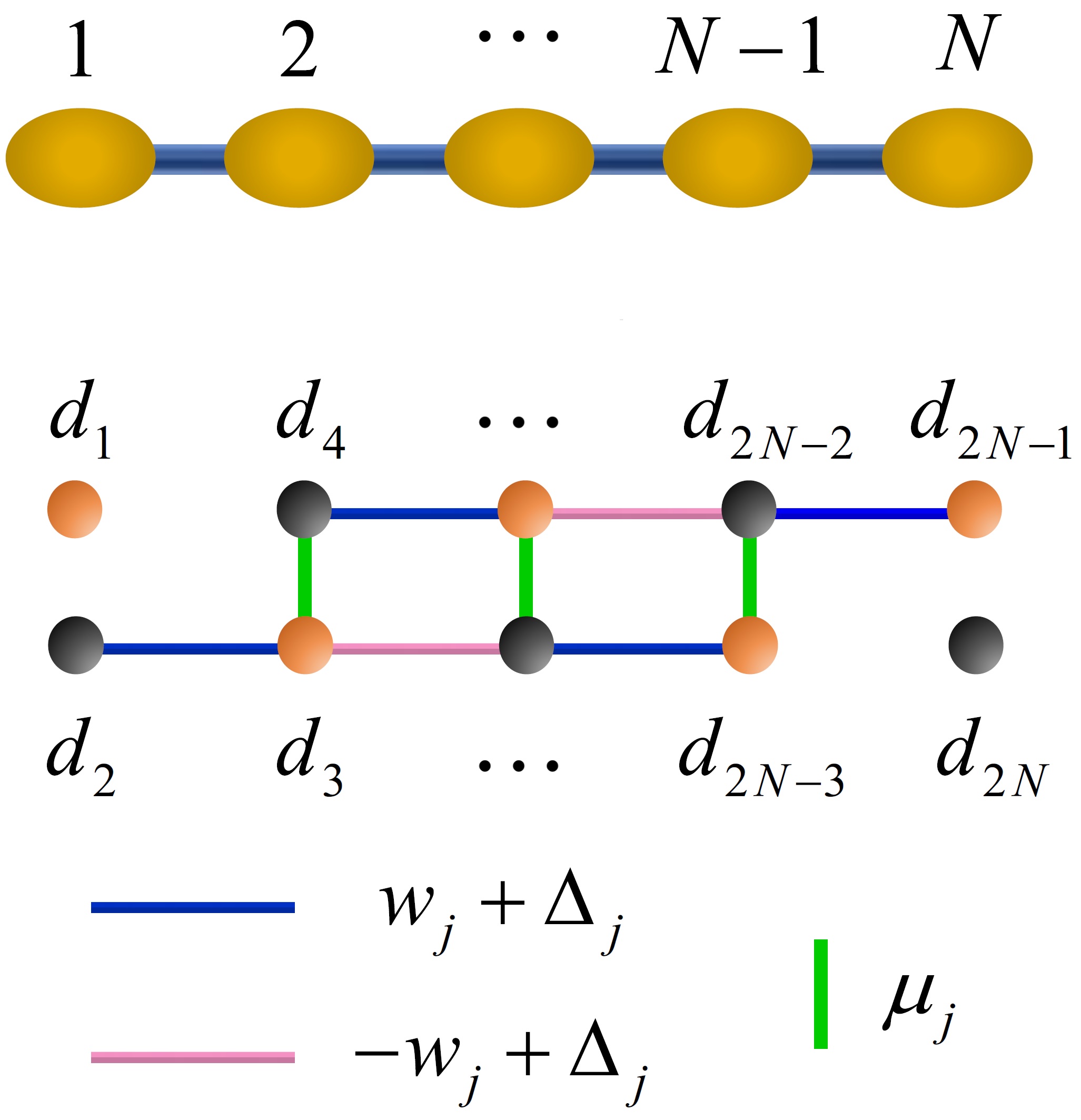}
\caption{\textbf{The Majorana representation of the chain. } A quantum-dot chain with $N$ fermionic sites ($N=5$) is represented by $2N$ MFs (each site by two MFs). The two boundary MFs are unpaired for the parameters in Eq.~(\ref{interesting-solution}). They form a zero-energy mode, while all the other MFs are fused to form $N-1$ new fermions. The lines of different colors connecting MFs represent the bonding strength according to Eq.~(\ref{m-chain}).}\label{Fig-4-MF-pairing}
\end{figure}

We assume $\lambda_{k}\neq 0$, so that the chain has only one zero-energy mode and the degeneracy of the ground states is two. It can be numerically verified that the assumption is valid when $\Delta_{j},w_{j}\neq 0$, $\Delta_{j}\approx w_{j}$ and $\mu_{j}\approx 0$ throughout the chain. The experimental realization of Eq.~(\ref{interesting-solution}) is feasible, as the chemical potential can be finely tuned through varying the gate voltage while the tunneling and pairing interactions can be adjusted through applying local tilted magnetic fields to the boundary sites\cite{two-dot-chain,nanoscale-B}. See also Ref.\cite{chain-other-experiments} for alternative implementations of a quantum-dot chain with different tuning methods.

The zero-energy mode $\tilde{b}_{N}^{\dagger}\tilde{b}_{N}$ corresponds to two-fold degenerate ground states with definite parity:
\ba\label{two-ground-states}
\ket{G_{1}}&=&\ket{\tilde{0}}_{1}\ket{\tilde{0}}_{2}\cdots\ket{\tilde{0}}_{N-1}\ket{\tilde{0}}_{N}\,,\nonumber\\
\ket{G_{2}}&=&\tilde{b}_{N}^{\dagger}\ket{G_{1}}=\ket{\tilde{0}}_{1}\ket{\tilde{0}}_{2}\cdots\ket{\tilde{0}}_{N-1}\ket{\tilde{1}}_{N}\,. \ea
They satisfy $\tilde{b}_{k}\ket{G_{1}}=\tilde{b}_{k}\ket{G_{2}}=0,(k=1,2,\cdots,N-1)$, $\tilde{b}_{N}\ket{G_{1}}=0$, and $\tilde{b}_{N}^{\dagger}\ket{G_{2}}=0$. In realistic situations, the chain of quantum dots is coupled to a reservoir that is composed of various sources such as phonons, superconducting grains, noisy electric gates, etc.  The reservoir may induce transitions between different quantum states of the chain. This is the phenomenon of quasi-particle poisoning\cite{quasi-particle-poisoning-0,quasi-particle-poisoning}. Suppose the perturbation from the reservoir is sufficiently weak as compared with the energy gap of the system. The effect of the perturbation will only be to induce transitions between the two ground states, which can cause bit-flip errors and decoherence in quantum computation when the qubits are encoded using these states. The density matrix of the chain can be written as
\ba\label{gs-mixture}
q\ket{G_{1}}\bra{G_{1}}+(1-q)\ket{G_{2}}\bra{G_{2}},
\ea
where $q$ depends on the details of the system-reservoir coupling mechanism ($0\leq q\leq1$). Note that the density matrix is diagonal in the basis of $\ket{G_{1}}$ and $\ket{G_{2}}$ because the superposition between the two basis vectors differing by fermionic parity is not allowed\cite{quasi-particle-poisoning}. One can use Josephson junctions to determine whether the ground state is $\ket{G_{1}}$ or $\ket{G_{2}}$ through measuring its parity\cite{MF-etg-1, MF-readout, MF-readout-2}.

By using $\tilde{b}_{N}\ket{G_{1}}=0$ and $\tilde{b}_{N}^{\dagger}\ket{G_{2}}=0$, it is not difficult to find out the forms of the two ground states in real space of the chain (the details are presented in Supplementary Materials).
\ba\label{ground-state-real-space-solution}
\ket{G_{1}}&=&x_{0}\ket{\Phi^{(\eta)}}_{1,N}\ket{S_{e}}+x_{1}\ket{\Psi^{(\eta)}}_{1,N}\ket{S_{o}},\nonumber\\
\ket{G_{2}}&=&x_{0}\ket{\Psi^{(\eta)}}_{1,N}\ket{S_{e}}+x_{1}\ket{\Phi^{(\eta)}}_{1,N}\ket{S_{o}},
\ea
where the states $\ket{\Phi^{(\eta)}}_{1,N}=(\ket{00}_{1,N}+\eta\ket{11}_{1,N})/\sqrt{2}$ and $\ket{\Psi^{(\eta)}}_{1,N}=(\ket{10}_{1,N}+\eta\ket{01}_{1,N})/\sqrt{2}$ are maximally entangled states between the boundary sites $1$ and $N$, $\ket{S_{e}}$  and $\ket{S_{o}}$ are some states in general forms for the sites from $2$ to $N-1$, and $x_{0}$, $x_{1}$ are coefficients to be determined ($|x_{0}|^{2}+|x_{1}|^{2}=1$). For $N=2$, there are no inner sites and the two degenerate ground states are reduced to $\ket{\Phi^{(\eta)}}_{1,N}$ and $\ket{\Psi^{(\eta)}}_{1,N}$, which has been considered in Ref.\cite{two-dot-chain}.
The four sub-chains in Fig.~1 will have four respective states in the form of Eq.~(\ref{gs-mixture}). We shall employ some schemes (discussed in subsequent subsections) to extract $\ket{\Phi^{(\eta)}}_{1,N}$ or $\ket{\Psi^{(\eta)}}_{1,N}$, and further to transform them into two useful entangled qubits. The parameter $\eta$ is determined by $w_{j},\Delta_{j},\mu_{j}$ in Eq.~(\ref{m-chain}). When $\eta=1$, the process of diagonalizing~(\ref{m-chain}) corresponds to a \emph{proper rotation} with the real orthogonal matrix $W_{0}$ acting on the vector $\vec{d}=(d_{2},d_{3},\cdots,d_{2N-1})$. The case of $\eta=-1$ is an \emph{improper rotation} which includes a reflection operation on $\vec{d}$. It is numerically found that when $\Delta_{j}\approx w_{j}>0$, $\mu_{j}\approx 0$, we have $\eta=1$. This is the parameter regime that we would like to consider, while the case $\eta=-1$ involves changing the signs of an odd number of $w_{j}$ and $\Delta_{j}$ (a reflection operation) which is not a typical situation in experiments. Therefore, without loss of generality, we shall assume $\eta=1$.

The ground states~(\ref{ground-state-real-space-solution}) are a topological phase of the chain. This is because their structure depends only on the boundary conditions~(\ref{interesting-solution}), not on the details of the inner sites, as in~(\ref{interesting-solution}), $w_{k},\Delta_{k},\mu_{k}$ in the bulk of the chain are arbitrary. However, their values will influence $x_{0},x_{1}$ and the details of $\ket{S_{e}},\ket{S_{o}}$ in~(\ref{ground-state-real-space-solution}), and also the energy gap ($\sim\textrm{Min}(2|\Delta_{j}|)$ on condition that $\Delta_{j}\approx w_{j}$ and $\mu_{j}\approx 0$ throughout the chain, see Ref.\cite{chain-experiment}). As discussed earlier, we assume the degeneracy of the ground states is always two, otherwise additional ground states may recombine with~(\ref{ground-state-real-space-solution}), thus changing their structure. For instance, consider two chains that each chain has a zero-energy mode with two degenerate ground states. So, there are four ground states if the two chains are viewed as a single chain. The states with the same parity can be freely transformed within their subspace, and~(\ref{ground-state-real-space-solution}) is one possible result but not the only one.

Finally, we shall show that the ground states are protected against perturbations that respect characteristic symmetries of the chain. The Hamiltonian~(\ref{Kitaev-chain}) in general has two symmetries\cite{MF-classify,chain-experiment}: the $\mathbb{Z}_2$ symmetry (the parity operator $P=\prod_{j=1}^{N}(1-2c_{j}^{\dagger}c_{j})$ commutes with $H$, $[P,H]=0$), and the particle-hole anti-symmetry ($H$ changes to $-H+constant$ when $c_{j}$ changes to $c^{\dagger}_{j}$). In addition, the system has another symmetry when the condition~(\ref{interesting-solution}) is fulfilled. We notice from Eq.~(\ref{m-chain}) and ~(\ref{interesting-solution}) that $[d_{1},H]=[d_{2N},H]=0$. In fact, $d_{1},d_{2N}$ realize the particle-hole transformation on the boundary sites: $d_{1}c_{1}d_{1}^{\dagger}=c_{1}^{\dagger}$, $d_{2N}c_{N}d_{2N}^{\dagger}=-c_{N}^{\dagger}$. Define a unitary operator $X=d_{1}d_{2N}$. We have $[X,P]=[X,H]=0$, $Xc_{1}X^{\dagger}=-c_{1}^{\dagger}$, $Xc_{N}X^{\dagger}=c_{N}^{\dagger}$, while $[X,c_{j}]=[X,c_{j}^{\dagger}]=0$ for $2\leq j\leq N-1$. Namely, $X$ realizes the particle-hole transformation on the two boundary sites simultaneously. Therefore, the eigenvalues of $P,X$ can be used to classify the eigenstates of $H$. In particular, $X\ket{G_{k}}=(-1)^{k+1}i\eta\ket{G_{k}}$, $k=1,2$. Consider a perturbation $H_{p}$ (static or time-dependent) whose energy scale is much smaller than the energy gap of the chain, so that the induced transition to excited states can be neglected and $H_{p}$ effectively only acts on the ground-state subspace\cite{topological-phase}. In this situation, either $[H_{p},X]=0$ or $[H_{p},P]=0$ will guarantee the stability of the ground states (namely $\bra{G_{2}}H_{p}\ket{G_{1}}=0$). 
In our system, $[H_{p},X]=0$ corresponds to $H_{p}$ in general not acting on the edges of the chain (unless $H_{p}$ involves only $d_{2},d_{2N-1},iX$ of the edges), which is a topological condition of $H_{p}$. Hence the protection of the ground states is said to be topological.  $[H_{p},P]=0$ corresponds to $H_{p}$ typically containing no hopping or paring interactions with environment, which offers an extra possibility of protecting the ground states when the topological condition of $H_{p}$ is not fulfilled.

In addition to the topological protection mentioned above, another interesting property of the system is that from Eq.~(\ref{gs-mixture}) and (\ref{ground-state-real-space-solution}) the reduced state of the two boundary sites can be calculated as
\ba\label{reduced-state-boundary}
r\ket{\Psi^{(\eta)}}_{1,N}\bra{\Psi^{(\eta)}}_{1,N}+(1-r)\ket{\Phi^{(\eta)}}_{1,N}\bra{\Phi^{(\eta)}}_{1,N}\,,
\ea
where $r=q+(1-2q)|x_{0}|^{2}$. It can be seen that the probability distribution $(r,1-r)$ of the two maximally entangled states depends on the system-reservoir coupling through $q$ and the inner part of the system through $x_{0}$. However, the states $\ket{\Psi^{(\eta)}}_{1,N}$ and $\ket{\Phi^{(\eta)}}_{1,N}$ with distinct parities are independent of these factors. This indicates that the states of the boundary sites with a definite parity are robust against quasi-particle poisoning as well as the disorders of the bulk, on condition that the perturbation from the reservoir is much smaller than the energy gap of the system. Note here that $\eta$ is fixed for a specific system, provided the same condition is fulfilled.

\subsection*{Measurement scheme.} The expression~(\ref{reduced-state-boundary}) motivates us to extract the entangled states of the boundary sites with a definite parity. The entangled states can be viewed as a charge qubit in an equal superposition of its two basis vectors $\ket{00}_{1,N}$ and $\ket{11}_{1,N}$\,, or $\ket{10}_{1,N}$ and $\ket{01}_{1,N}$. We shall show in the next subsection that four such qubits can be used to prepare two entangled qubits. At this juncture, we would like to remark that these qubits are different from the topological qubits which are encoded in the degenerate ground state subspace of two chains (i.e. the encoding basis vectors are $\ket{G_{1}}\ket{G_{1}}$ and $\ket{G_{2}}\ket{G_{2}}$\,, or $\ket{G_{1}}\ket{G_{2}}$ and $\ket{G_{2}}\ket{G_{1}}$, cf. Eq.~(\ref{two-ground-states})). Although our qubits are no longer topologically protected against the environmental noise, there are still advantages over the topological qubits. For example, the topological qubit is susceptible to the quasi-particle poisoning\cite{quasi-particle-poisoning-0,quasi-particle-poisoning} which induces transitions between $\ket{G_{1}}$ and $\ket{G_{2}}$ causing bit-flip errors and decoherence in quantum computation. Suppose the topological qubit is in the state $\ket{\psi_{q}}=(\ket{G_{1}}\ket{G_{2}}+\ket{G_{2}}\ket{G_{1}})/\sqrt{2}$ encoded in two chains. The reservoir is simulated using a minimal model\cite{quasi-particle-poisoning-0}: an additional fermionic site in the vacuum state $\ket{0}$ and coupled to the first chain with the Hamiltonian\cite{quasi-particle-poisoning-0,reservoir-tunnel} $H_{d}=\epsilon c^{\dagger}c+\kappa(c^{\dagger}-c)(\tilde{b}^{\dagger}+\tilde{b})$, where $c^{\dagger},c$ ($\tilde{b}^{\dagger},\tilde{b}$) are the creation and annihilation operators of the fermionic site (the zero-energy edge mode of the first chain), and $\epsilon$ is the energy of the fermion. With the time evolution $e^{-itH_{d}}\ket{0}\ket{\psi_{q}}$, the topological qubit will be entangled with the reservoir, which destroys the coherence of the qubit (for special parameters e.g. $\epsilon=0,\,t=\pi/2\kappa$ the qubit gets disentangled with the reservoir with a flip in half of the encoding basis: $\ket{G_{1}}$ and $\ket{G_{2}}$ of the first chain are interchanged). In contrast, our qubit before extraction is in the state (\ref{reduced-state-boundary}) for which the quasi-particle poisoning only affects the probability distribution of the two types of encoding for the qubit ($\ket{\Psi^{(\eta)}}_{1,N}$ or $\ket{\Phi^{(\eta)}}_{1,N}$), while the coherence of the qubit (i.e. the superposition between the encoding basis vectors) remains intact. Therefore, the state (\ref{reduced-state-boundary}) can be regarded as a quantum memory for preserving the qubits (or entanglement) encoded in the two boundary sites. As will be shown later in the present subsection,  during the extraction the boundary sites are isolated from the environment (except the microwave) through increasing the confining potential for them. Thus the quasi-particle poisoning is not an issue in this process.

To extract the entangled states of the
boundary sites, one could first determine the ground state through measuring its parity\cite{MF-etg-1, MF-readout, MF-readout-2}. This is sufficient to achieve the extraction for $N=2$. For $N>2$, it is necessary to further measure the parity of all the inner-site electrons in order to collapse the ground state~(\ref{ground-state-real-space-solution}) into a configuration that the boundary sites are either in $\ket{\Psi^{(\eta)}}_{1,N}$ or in $\ket{\Phi^{(\eta)}}_{1,N}$. This measurement can be performed by using single-electron detectors to directly probing the electron in each of the inner sites. The summation of all the measurement results ($0$ or $1$ for each inner site) gives an even or odd number representing the parity of all the inner-site electrons. However, the method requires the inner sites to be decoupled from each other, otherwise the detector will only couple to the eigen modes of the chain and fail to measure the electron of the individual sites. The requirement can be fulfilled through increasing the confining potential for each inner site, which is complex for a long chain. A simpler way is to directly measure the parity of the boundary-site electrons. The measurement should be coherent, not by counting the electrons in the boundary sites. Namely the reduced state~(\ref{reduced-state-boundary}) should be collapsed to one of its two terms when the measurement is done. To this end, one could couple a microwave dispersively to the boundary sites in order to measure their parity. There are three ways of realizing the dispersive coupling. (1). A microwave cavity (the transmission line resonator, TLR) can be designed with protrusions\cite{cavity-protrusion-0,cavity-protrusion} in the region of the boundary sites of the sub-chain in order to concentrate the electromagnetic field locally, as shown in Fig.~5. The interaction of the microwave with the inner part of the sub-chain is neglected. (2). A local magnetic field\cite{nanoscale-B} can be applied adiabatically to the boundary sites, in order to increase the energy gap between the spin-split levels of the boundary-site electrons which will be off-resonant with the inner part of the sub-chain. In this way, the microwave, when applied to the entire sub-chain, will interact only effectively with the boundary sites (i.e. the microwave is off-resonant with and thus decoupled from the inner part of the sub-chain). (3). The quantum state of each boundary site can be transferred to an ancillary site through hopping of electrons ($e^{-itH}$, $H=-w(c^{\dagger}_{1}c_{2}+c^{\dagger}_{2}c_{1})$, $t=\frac{\pi}{2w}$). The ancillary sites are not aligned with the chain, and the parity measurement will be performed on them. They can even form the new boundaries of the original chain, with the overall state described by Eq.~(\ref{gs-mixture}). The advantage of the third scheme is that the interaction of the microwave with the inner part of the chain is completely eliminated. Here we only discuss the first scheme. Apparently, the discussion when slightly modified applies to the other two schemes as well.

\begin{figure}
\centering
\includegraphics[width=0.45\textwidth,height=0.27\textwidth]{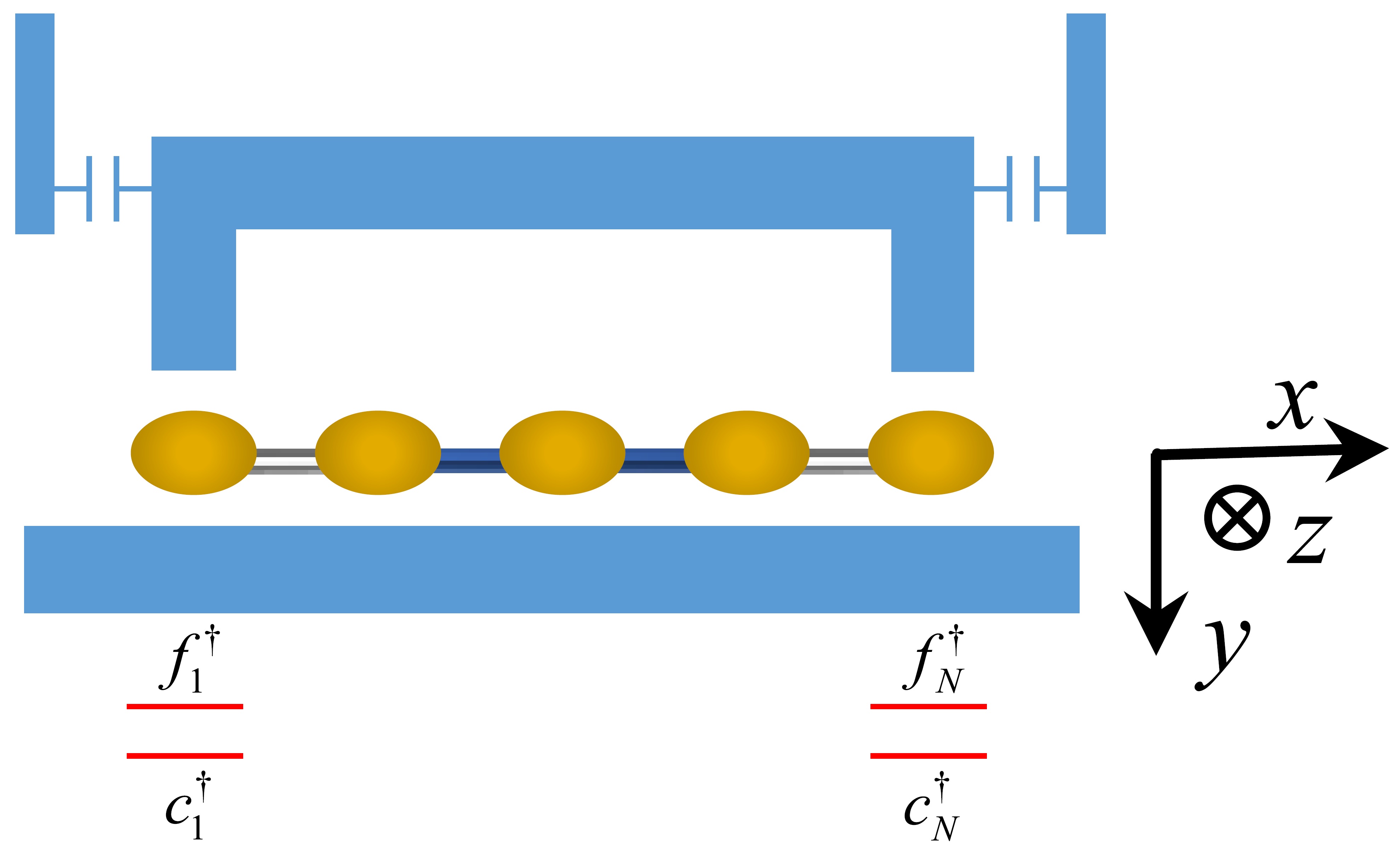}
\caption{\textbf{The schematic diagram of the parity measurement. }A microwave transmission line resonator is employed to measure the parity of the boundary-site electrons of a sub-chain, with $f_{j}^{\dagger}/c_{j}^{\dagger}$ ($j=1,N$) the creation operator of the upper/lower spin-split energy level (red line) of the boundary quantum dots. Here the interaction of the boundary sites with the inner part of the chain has already been switched off through increasing the confining potential for the boundary sites. The gemometry of the setup allows the electromagnetic field to be concentrated on the boundary sites, and its interaction with the inner part of the chain is neglected. See also Ref.\cite{cavity-protrusion-0,cavity-protrusion} for similar geometric designs where plunger gates/protrusions are used to concentrate the electromagnetic field. The accuracy of this scheme can be enhanced through applying local magnetic fields to make the boundary sites off-resonant with the inner part of the chain.}\label{Fig-5-parity-measure}
\end{figure}

We shall show that after a proper time of interaction, the boundary sites initially in the state~(\ref{reduced-state-boundary}) and TLR initially in the coherent state $\ket{\alpha}$ will evolve to
\ba\label{parity-measure-four-coherent}
r\ket{\Psi^{(\eta)}}_{1,N}\bra{\Psi^{(\eta)}}_{1,N}\otimes\ket{-\alpha}\bra{-\alpha}+(1-r)\ket{\Phi^{(\eta)}}_{1,N}\bra{\Phi^{(\eta)}}_{1,N}\otimes\ket{\alpha}\bra{\alpha},
\ea
where $\ket{\pm\alpha}$ are the coherent states of TLR with the respective amplitude $\pm\alpha$, and other symbols are those in Eq.~(\ref{reduced-state-boundary}). It can be seen that the state of the boundary sites will collapse to $\ket{\Phi^{(\eta)}}_{1,N}$ or $\ket{\Psi^{(\eta)}}_{1,N}$ through measuring the coherent states of the microwave, provided the coherent states are orthogonal. The orthogonality can be achieved to a good approximation when the amplitude $\alpha$ is sufficiently large so that $|\langle\alpha|-\alpha\rangle|^{2}=e^{-4|\alpha|^{2}}\ll 1$. The two coherent states can be measured through homodyne detection by means of parametric amplifiers and mixers\cite{U-dot-photon-phase-int}.

The remaining part of the subsection will concentrate on the derivation of Eq.~(\ref{parity-measure-four-coherent}), and based on it another measurement scheme is mentioned in the last paragraph. First, we suddenly increase the strength of the confining potential in the direction along the chain for the boundary sites so that they are decoupled from the inner sites. This is a process of quantum quench. The effective Hamiltonian (including the measurement setup in Fig.~5) with rotating-wave approximation is
\ba\label{microwave-dot-H}
H_{in}=\frac{\delta}{2}(f_{1}^{\dagger}f_{1}+f_{N}^{\dagger}f_{N})+Je^{i\phi}a(f_{1}^{\dagger}c_{1}+f_{N}^{\dagger}c_{N})+h.c.,
\ea
where $\delta=\Omega-\omega_{m}$ is the detuning between the resonant frequency ($\Omega$) of the quantum dots (the energy gap of the spin-split levels) and the frequency ($\omega_{m}$) of the microwave photons, $c_{j},c_{j}^{\dagger}$ are defined in Eq.~(\ref{Kitaev-chain}), $f_{j} (f_{j}^{\dagger})$ is the annihilation (creation) operator for the upper spin-split level of the $j$th quantum dot around the Fermi level (see Fig.~5), $a^{\dagger},a$ are the creation and annihilation operators of the microwave photons,
$J$ is the spin-photon coupling strength induced by the spin-orbit interaction in the quantum dots\cite{photon-dot-interact-2} (see Supplementary Materials for the estimation of $J$), $J/\delta\ll 1$ (the dispersive coupling regime), $h.c.$ denotes the hermitian conjugation of its previous terms, and $\phi$ is a tunable phase. The Hamiltonian is written in the interaction picture with the free part $H_{0}=\omega_{m}(a^{\dagger}a+\frac{1}{2}+f_{1}^{\dagger}f_{1}+f_{N}^{\dagger}f_{N})$, and the chemical potentials of the boundary sites are finely tuned to zero.

The Hamiltonian~(\ref{microwave-dot-H}) essentially describes the interaction between photons and two-level atoms, if we define the raising and lowering operators for the two-level atoms as $\sigma_{j}^{\dagger}=f_{j}^{\dagger}c_{j}$, $\sigma_{j}=c_{j}^{\dagger}f_{j}$. The anti-commutator $\sigma_{j}^{\dagger}\sigma_{j}+\sigma_{j}\sigma_{j}^{\dagger}=f_{j}^{\dagger}f_{j}+c_{j}^{\dagger}c_{j}-2f_{j}^{\dagger}f_{j}c_{j}^{\dagger}c_{j}=1$ is fulfilled when there is exactly one electron in the two spin-split levels around the Fermi level. However, when there is no electron around the Fermi level of some boundary site, the above anti-commutator equals $0$ and thus is not well-defined. In this situation, the microwave will not effectively interact with that specific site. Therefore, the number of the two-level atoms (denoted as $N_{0}$) in the essential TC model is not fixed; it depends on the number of electrons around the Fermi levels of boundary sites. The parity of this number is what we want to measure.

Next we apply a unitary transformation $U=\exp[\frac{J}{\delta}\sum_{j}(a\sigma_{j}^{\dagger}-a^{\dagger}\sigma_{j})]$ to the Hamiltonian~(\ref{microwave-dot-H}) for $\phi=0$. Expanding to second order in $J/\delta$, we have\cite{U-dot-photon-phase-int,time-evolution-parity}
\ba\label{U-transform}
H_{in}'=UH_{in}U^{\dagger}\approx(\delta+\frac{J^{2}}{\delta})\sum_{j}\sigma_{j}^{\dagger}\sigma_{j}+\frac{J^{2}}{\delta}\sum_{j}\sigma_{j}^{z}a^{\dagger}a+\frac{J^{2}}{\delta}\sum_{j\neq k}(\sigma_{j}^{\dagger}\sigma_{k}+\sigma_{j}\sigma_{k}^{\dagger}),\hspace{0.8cm}
\ea
where $\sigma_{j}^{z}=\sigma_{j}^{\dagger}\sigma_{j}-\sigma_{j}\sigma_{j}^{\dagger}$. The terms proportional to $J^{3}$ are neglected for an initial coherent state of the microwave with its amplitude $\alpha$ satisfying $\alpha\ll\delta/(2J)$.

We notice that the initial state of the $N_{0}$ atoms is always in the lower spin-split levels, so that the last summation in Eq.~(\ref{U-transform}) is essentially $0$. For the microwave initially in a coherent state $\ket{\alpha}$ and the number of the boundary-site electrons to be $N_{0}$ (in a number state $\ket{N_{0}}$ created by $c_{j}^{\dagger}$'s), the state for a time evolution $T=\pi\delta/J^{2}$ becomes $e^{-iTH_{in}'}\ket{\alpha}\ket{N_{0}}=\ket{(-1)^{N_{0}}\alpha}\ket{N_{0}}$, where $\ket{(-1)^{N_{0}}\alpha}$ is a coherent state of the microwave with amplitude $(-1)^{N_{0}}\alpha$. It can be seen that this result is consistent with Eq.~(\ref{parity-measure-four-coherent}). We also notice that $U,U^{\dagger}$ in Eq.~(\ref{U-transform}) can be realized through adiabatically tuning the gap of the boundary-site electrons into resonance with the microwave and then adjusting the phase $\phi$ to be $\pi/2, -\pi/2$ respectively for a time evolution $t=1/\delta$ (see Eq.~(\ref{microwave-dot-H}) and Fig.~3(a)). The microwave is stored in a quantum memory\cite{quantum-memory-1,quantum-memory-2} in the process of tuning the gap of the boundary-site electrons. Hence the derivation is finished. Experimentally\cite{chain-experiment}, $\omega_{m}/(2\pi)\sim114.31$ GHz, $\Omega/(2\pi)\sim120.74$ GHz ($0.5$ meV), $J/(2\pi)\sim214$ MHz, $\alpha\sim1.5$ and $T\sim70$ ns. If we choose $T=\pi\delta/(2J^{2})$ instead of $\pi\delta/J^{2}$, the state will evolve to $\ket{(i)^{N_{0}}\alpha}\ket{N_{0}}$ which can still be used to measure the parity of $N_{0}$. However, despite the advantage of shorter time, measuring the corresponding four coherent states will involve higher error rates as the overlap among them increases.

Another method for measuring the parity of the boundary-site electrons is through measuring the transmission spectrum of TLR\cite{U-dot-photon-phase-int}. As can been seen from Eq.~(\ref{U-transform}), the TLR frequency is shifted by $(J^{2}/\delta)\sum_{j}\sigma_{j}^{z}$ which depends on the state of the boundary sites. The shift is $-J^{2}/\delta$ for $\ket{\Psi^{(\eta)}}_{1,N}$ and $-2J^{2}/\delta$ for $\ket{11}_{1,N}$ (no shift for $\ket{00}_{1,N}$). If we drive TLR at the frequency $\omega_{m}-J^{2}/\delta$, the photon will be transmitted for $\ket{\Psi^{(\eta)}}_{1,N}$ and reflected for $\ket{\Phi^{(\eta)}}_{1,N}$. To make this method accurate, the photon loss of TLR, which causes spectral line broadening, need be reduced. 

\subsection*{Useful entanglement resource.} The entangled states extracted from (\ref{reduced-state-boundary}) are not a useful entanglement resource: e.g. one can neither test Bell inequalities\cite{CHSH} nor perform quantum computing\cite{Teleport-QC} with these states. This is because the two levels of one subsystem of the entangled states are represented by the absence ($\ket{0}$) and presence ($\ket{1}$) of a fermion in a single site. There is no physical mechanism that could be used to prepare a superposition of the two levels differing by fermionic parity: $\alpha\ket{0}+\beta\ket{1}$, ($|\alpha|^2+|\beta|^2=1$). It's required to use two fermionic sites\cite{qubit-parafermion,DQD-QC}, e.g. $\ket{\overline{0}}\equiv\ket{10},\ket{\overline{1}}\equiv\ket{01}$.  This is similar to the encoding of topological qubits. Indeed, we have already called the extracted entangled state a charge qubit rather than two entangled qubits in the previous subsection. An interesting question is whether or not it is possible to prepare two entangled qubits. This can be done through the following scheme. Suppose we have extracted an entangled state from each sub-chain shown in Fig. 2(a) and all the states are $\ket{\Psi^{(\eta)}}_{1,N}$ with $\eta=1$. Let us focus on the sub-chain $C_{1}$ and $C_{3}$ first. The state of the sites $1,2,3,4$ in Fig. 2(c) is
\ba\label{initial-state-etg-qubit}
\ket{\psi}_{1234}=\frac{\ket{10}+\ket{01}}{\sqrt{2}}\frac{\ket{10}+\ket{01}}{\sqrt{2}}
\ea
Then, we swap the states of the site $2$ and $3$ through the time evolution with the Hamiltonian $H_{23}=-w(c_{2}^{\dagger}c_{3}+c_{3}^{\dagger}c_{2})$. Note that $\exp(-itH_{23})=\ket{00}\bra{00}+\ket{11}\bra{11}+i\ket{01}\bra{10}+i\ket{10}\bra{01}$ when $t=\frac{\pi}{2w}$. So we have
\ba\label{logical-etg}
e^{-itH_{23}}\ket{\psi}_{1234}=\frac{1}{\sqrt{2}}[\ket{10}_{13}(\frac{i\ket{10}_{24}+\ket{01}_{24}}{\sqrt{2}})+\ket{01}_{13}(\frac{\ket{10}_{24}+i\ket{01}_{24}}{\sqrt{2}})],
\ea
where we have grouped the states of the site $1$ and $3$ together (and also for $2,4$). The swap operation in Eq.~(\ref{logical-etg}), involving a phase factor $i$, is referred to as the $i$-swap. This is different from the ideal exchange operation (without the phase $i$), which was used e.g. in the spin system\cite{swap-HBT} to simulate the Hanbury Brown-Twiss Interferometer in quantum optics (the $i$-swap can also be used but it is dispensable there). It can be seen that we have obtained a maximally entangled state with two qubits encoded by four fermionic sites: the sites $1,3$ for one qubit and $2,4$ for the other. The logical basis are $\ket{\overline{0}}_{1}\equiv\ket{10}_{13}$, $\ket{\overline{1}}_{1}\equiv\ket{01}_{13}$ for the first qubit, and $\ket{\overline{0}}_{2}\equiv\ket{10}_{24}$, $\ket{\overline{1}}_{2}\equiv\ket{01}_{24}$ for the second. In fact, the success of creating the entangled state is attributed to the phase factor $i$ mentioned earlier; without it the above process is not possible\cite{fermion-swap-graph-state}. See Fig.~6 for the schematic diagram of the swap operation. One drawback of the entangled qubits is that the site $3$ is far from the site $1$ so that it is difficult to induce interaction between them (similar for the site $2$ and $4$). This problem can be solved by using the chain $C_{1'}$ and $C_{3'}$ as shown in Fig.~2 (a). Both of the chains are in the maximally entangled states as $C_{1}$ and $C_{3}$. Then, one can perform teleportation to transfer the state of the site $3$ to $1'$ and that of $2$ to $4'$. The detailed scheme is discussed in Supplementary Materials. Finally, we obtain a useful maximally entangled state with two logical qubits encoded by the sites $1,1'$ and $4,4'$ respectively. See Fig.~2(d).

\begin{figure}
\centering
\includegraphics[width=0.4\textwidth,height=0.2\textwidth]{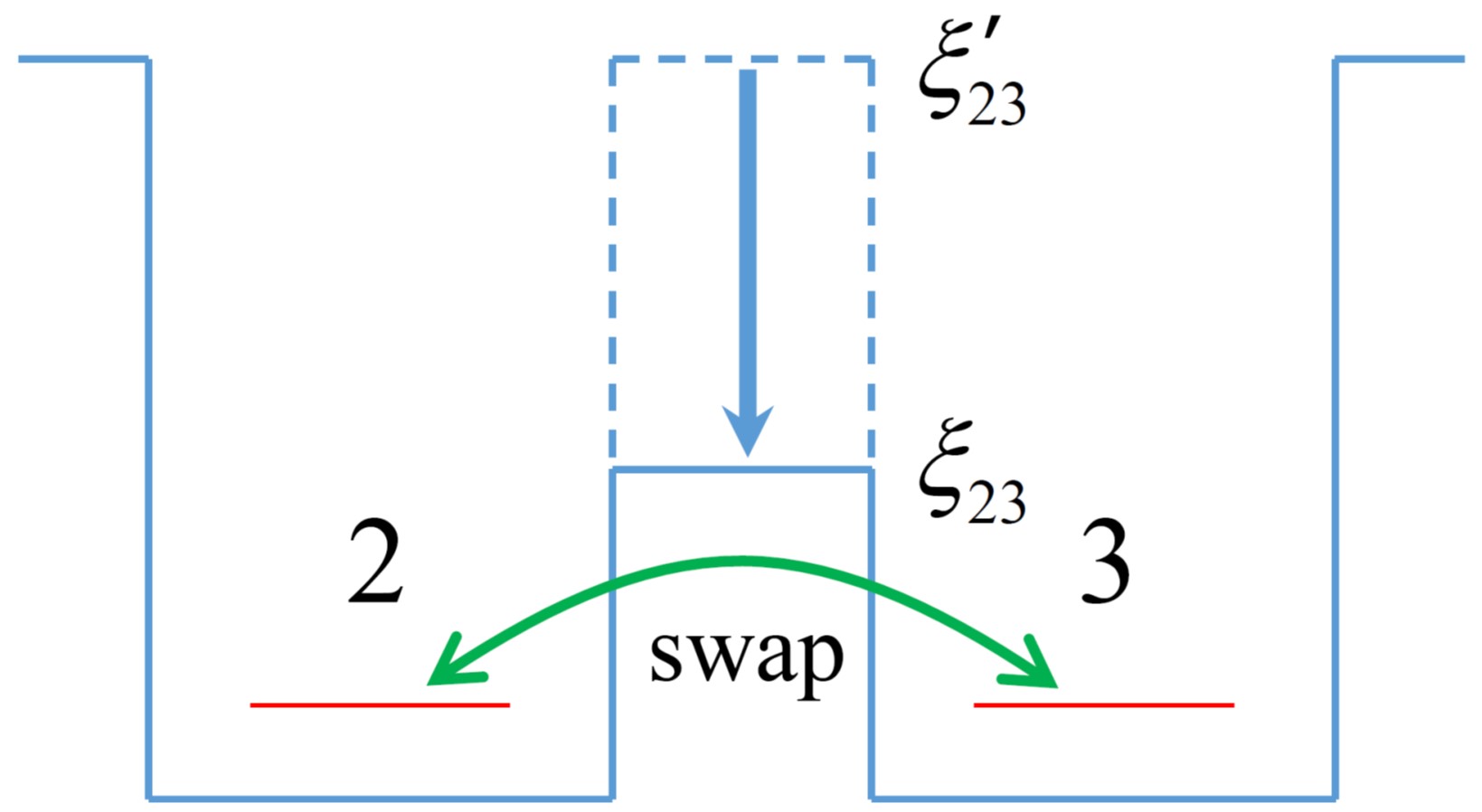}
\caption{\textbf{The schematic diagram of the swap operation. }$\xi_{23}'$ is the potential barrier between the quantum dot $2$ and $3$ (see Fig.~2(c)). The barrier is sufficiently high so that the coupling between the two quantum dots is switched off. When the barrier is suddenly decreased from $\xi_{23}'$ to $\xi_{23}$, the hopping interaction between the two sites is induced (see the green line with arrows). This interaction results in the swap operation as shown in Eq.~(\ref{logical-etg}). The decrease of the barrier is realized through increasing the gate voltage between the two sites. See Fig.~3(b). After the swap operation is finished, the barrier is restored to the higher value $\xi_{23}'$ in order to switch off the interaction between the two sites.}\label{Fig-6-swap}
\end{figure}

The above scheme is still feasible when the measurement result for the parity of the boundary-site electrons of $C_{1}$ and/or $C_{3}$ is even, with a little difference in the final entangled state. For instance, if the parity of the boundary sites of $C_{1}$ is even, the logical basis for the qubit in the sites $1,1'$ will be $\ket{\overline{0}}_{1}\equiv\ket{00}_{11'}$, $\ket{\overline{1}}_{1}\equiv\ket{11}_{11'}$. In fact, the local gates for correcting the bit-flip errors or phase errors in the teleportation is dispensable for achieving the final entangled qubits. For instance, without performing the bit-flip gates, the basis of the relevant qubit will change from $\{\ket{00},\ket{11}\}$ to $\{\ket{10},\ket{01}\}$ (or reversely), but the entanglement is equivalent.

\noindent\textbf{Decoherence. }In realistic experiments, there are photon losses, and the quantum dots have dephasing due to the fluctuations of the electrical gate bias, in addition to the spontaneous emission. We use the master-equation approach to simulate the decoherence process.
\ba\label{master-eq}
\frac{d\rho}{dt}=&-&i[H_{in},\rho]+\kappa D[a]\rho+\frac{\gamma_{\phi_{1}}}{2}D[\sigma_{1}^{z}]\rho+\kappa_{1}(\bar{n}_{1}+1)D[\sigma_{1}]\rho+\kappa_{1}\bar{n}_{1}D[\sigma_{1}^{\dagger}]\rho\nonumber\\
&+&\frac{\gamma_{\phi_{N}}}{2}D[\sigma_{N}^{z}]\rho+\kappa_{N}(\bar{n}_{N}+1)D[\sigma_{N}]\rho+\kappa_{N}\bar{n}_{N}D[\sigma_{N}^{\dagger}]\rho,
\ea
where $H_{in}$ is defined in Eq.~(\ref{microwave-dot-H}), $D[\mathcal{O}]\rho=(2\mathcal{O}\rho\mathcal{O}^{\dagger}-\mathcal{O}^{\dagger}\mathcal{O}\rho-\rho\mathcal{O}^{\dagger}\mathcal{O})/2$, $\kappa$ is the photon decay rate of TLR, $\gamma_{\phi_{j}}$ and $\gamma_{j}$ are the dephasing rate and spontaneous emission rate of the boundary quantum dots ($j=1,N$), and $n_{j}=(e^{\hbar\Omega/k_{B}T_{R}}-1)^{-1}$ is the mean photon number at the reservoir temperature $T_{R}$ and the transition frequency $\Omega$ between the spin-split levels of the $j$th quantum dot. Note that we have ignored the decoherence during the time evolution for realizing $U^{\dagger},U$ in Eq.~(\ref{U-transform}), as their operation time $\sim30$ ps, much smaller than the typical decoherence time ($\sim1\mu$s). Assume $\kappa/2\pi\sim 1$ MHz, $\gamma_{j}/2\pi\sim 0.2$ MHz, $\gamma_{\phi_{j}}/2\pi\sim 0.5$ MHz\cite{decoherence-values}, $T_{R}=10$ mK, and we consider the time evolution according to Eq.~(\ref{master-eq}). We calculate the fidelity of the reduced state of the two boundary sites with the ideal state $\ket{\Phi^{(\eta)}}_{1,N}$ or $\ket{\Psi^{(\eta)}}_{1,N}$, when the microwave's state is measured ($\ket{\pm\alpha}$). The fidelity between two states $\rho$ and $\tau$ is defined\cite{single-gates} as $F(\rho,\tau)=\textrm{tr}\sqrt{\rho^{1/2}\tau\rho^{1/2}}$. The numerical simulation shows that the better fidelity $0.97$ is obtained when the microwave's state is measured to be $\ket{-\alpha}$ (the other fidelity is $0.84$ for the microwave's state to be $\ket{\alpha}$; the large difference between the two fidelities is attributed to the fact that the state $\ket{\Psi^{(\eta)}}_{1,N}$ is less susceptible to the symmetric phase errors than $\ket{\Phi^{(\eta)}}_{1,N}$). The result indicates that our measurement scheme is feasible under realistic experimental conditions. If we ignore the further decoherence during the time evolution in Eq.~(\ref{logical-etg}), as it's much shorter ($\sim2$ ps) than the decoherence time ($\sim1\mu$s), the final fidelity for the entangled qubits is calculated to be as high as $0.9$.

\section*{Discussion}
We have shown that a pair of maximally entangled solid-state charge qubits can be extracted from two parallel chains of coupled quantum dots that support zero-energy edge modes. The edge mode is composed of unpaired Majorana fermions. The extracted entanglement is a useful resource for quantum computing\cite{Teleport-QC}. The details of the entangled state depends on both the parity measurement of the boundary-site electrons in each chain (see Eq.~(\ref{reduced-state-boundary})) and the teleportation between the chains. Our scheme provides an illustration of \emph{localizable entanglement}\cite{LE-2}, which is feasible under realistic experimental conditions, as it allows for finite temperature effect and local noise. That is, the ground states of the chain, from which the entangled charge qubits are extracted, are protected against thermal excitations due to the substantial energy gap, and they are also protected against perturbations that either respect the particle-hole symmetry of the boundary sites (a topological protection) or respect the $\mathbb{Z}_{2}$ symmetry of the chain. Even if the perturbation does not respect the symmetries of the system, it still cannot affect the extraction of the entangled qubits, as long as the perturbation is much smaller than the energy gap of the system. Namely, our scheme is robust against quasi-particle poisoning\cite{quasi-particle-poisoning-0,quasi-particle-poisoning}. In addition, it is only required to finely tune the parameters close to the boundary sites (see Eq.~(\ref{interesting-solution})), while small disorders in the inner sites (the bulk of the chain) is allowed.

The fidelity for the entangle qubits can be as high as $0.9$ for the decoherence that is mainly due to the noise of the electrical gate bias. As fluctuations of the environmental charges trapped in the insulating substrate or at the interface of the heterostructure may further reduce the coherence time\cite{dephasing-vs-trap}, new growth methods for materials with low trapped charge density\cite{low-trapped-charge}, as well as the charge echo techniques\cite{charge-echo}, can alleviate the decoherence. Our scheme, when combined with the purification protocols\cite{quantum-repeater-purify}, serves as a quantum repeater for distributing entanglement. The distance of the distribution are restricted mainly by the noise which is required to be much smaller than the energy gap of the system. The total noise grows linearly with the distance assuming independent and identical reservoirs for individual sites. The growth is less severe than the scheme for distributing entanglement through quantum state transfer, where the noise grows exponentially with the distance\cite{etg-transfer-1,etg-transfer-2}. Furthermore, due to the advantage (over the implementation using quantum wires) that the coherence length of the electrons are only required to be longer than the width of the individual superconducting grains between the nearest-neighbour quantum dots\cite{chain-experiment}, the distance of the distribution can be increased promisingly to macroscopic scale ($\gtrsim 10^{-4}$ m). This scale is compatible with the wavelength of the microwave ($\sim3$ mm) for realizing the parity measurement. Different from the existing proposals of quantum repeaters where postselection of photon states are used\cite{quantum-repeater-1,quantum-repeater-2}, our proposal involves manipulations of MFs (edge states).

Our proposal provides an experimental method for probing the structure of the topological ground states and might enable us to simplify the schemes of topological quantum computation using MFs. For instance, ancilla MFs are needed for realizing the topological two-qubit entangling gate, unless the four-MF interaction is realizable\cite{fermionic-QC}. But the corresponding edge states can be used for extracting conventional entangled qubits through the swap of Dirac fermions in our scheme (see Fig.~2(c)) without the aid of ancilla MFs. This difference can be further investigated, which helps to devise a way of removing the need for ancilla MFs in the process of performing the topological two-qubit entangling gate. In addition, future work can be pursued on the scenarios that the disorders close to the boundary sites are present and/or the disorders in the bulk are too large, in order to determine the range of parameters for a stable edge mode and the corresponding entangled qubits to exist.

\newpage

\section*{\huge Supplementary materials for ``Extracting entangled qubits from Majorana fermions in quantum dot chains through the measurement of parity''}

The supplementary materials consist of four sections. Sec. 1 discusses Bogoliubov-de Gennes transformation. Sec. 2 discusses the derivation of the ground states in real space of the chain. Sec. 3 gives an estimation of the electron-microwave coupling strength $J$. Sec. 4 discusses the teleportation for obtaining the final entangled qubits.

\section*{Bogoliubov-de Gennes transformation}\label{appendix-BdG-Eq}
One can use the Bogoliubov-de Gennes transformation\cite{BdG-transform} to solve for the zero mode
\ba\label{BdG-transform}
\tilde{b}_{j}=\frac{1}{2}\sum_{k=1}^{N}\left[\phi_{j,k}d_{2k-1}+i\psi_{j,k}d_{2k}\right]=\frac{1}{2}\sum_{k=1}^{N}\left[(\phi_{j,k}-\psi_{j,k})c_{k}^{\dagger}+(\phi_{j,k}+\psi_{j,k})c_{k}\right],\nonumber\\
(j=1,2,\cdots,N),
\ea
where $\phi_{j,k}$ and $\psi_{j,k}$ are real coefficients determined by the condition that the Hamiltonian (1) in the main text is diagonalized: $H=\sum_{j=1}^{N}\lambda_{j}(\tilde{b}_{j}^{\dagger}\tilde{b}_{j}-\frac{1}{2})$. We note that $[\tilde{b}_{j},H]=\lambda_{j}\tilde{b}_{j}$. If there exists a zero mode, say $\lambda_{N}=0$, we have $[\tilde{b}_{N},H]=0$. $[\tilde{b}_{N},H]$ can be calculated by using Eq.~(1) and~(\ref{BdG-transform}).
By requiring the coefficients of $c_{j}$ and $c_{j}^{\dagger}$ in the calculated result to be zero, we get
\ba\label{transfer-phi-psi}
\left(\begin{array}{c}
\phi_{N,j+1}\\
\phi_{N,j}
\end{array}\right)=&\left(\begin{array}{cc}
\frac{-\mu_{j}}{\Delta_{j}+w_{j}}&\frac{\Delta_{j-1}-w_{j-1}}{\Delta_{j}+w_{j}}\\
1&0
\end{array}\right)\left(\begin{array}{c}
\phi_{N,j}\\
\phi_{N,j-1}
\end{array}\right),\hspace{0.6cm}\nonumber\\
\left(\begin{array}{c}
\psi_{N,j-1}\\
\psi_{N,j}
\end{array}\right)=&\left(\begin{array}{cc}
\frac{-\mu_{j}}{\Delta_{j-1}+w_{j-1}}&\frac{\Delta_{j}-w_{j}}{\Delta_{j-1}+w_{j-1}}\\
1&0
\end{array}\right)\left(\begin{array}{c}
\psi_{N,j}\\
\psi_{N,j+1}
\end{array}\right),\nonumber\\
&(j=1,2,\cdots,N),\hspace{3.5cm}
\ea
where the variables with subscripts equal to $(N,0)$ or $(N,N+1)$ are assumed to be $0$. For a uniform chain i.e. $w_{j}=w$, $\Delta_{j}=\Delta$ and $\mu_{j}=\mu$ in the above two equations, we notice that the transfer matrices are identical: $\left(\begin{array}{cc}
\frac{-\mu}{\Delta+w}&\frac{\Delta-w}{\Delta+w}\\
1&0
\end{array}\right)
\equiv A$. However, the index for $\phi_{N,i}$ will increase upon the action of the transfer matrix, while for $\psi_{N,i}$ the index will decrease. So, we can set $\psi_{N,j}=\phi_{N,N+1-j}$, and the Bogoliubov-de Gennes transformation~(\ref{BdG-transform}) for $\tilde{b}_{N}$ becomes
\ba\label{BdG-transform-2-symmetry}
\tilde{b}_{N}&=&\frac{1}{2}\sum_{k=1}^{N}\left[\phi_{N,k}(d_{2k-1}+id_{2(N+1-k)})\right],\nonumber\\
&=&\frac{1}{2}\sum_{k=1}^{N}\left[\phi_{N,k}(c_{k}+c_{k}^{\dagger}+c_{N+1-k}-c_{N+1-k}^{\dagger})\right]
\ea
Suppose the two eigenvalues of the transfer matrix $A$ are $\lambda_{1}$ and $\lambda_{2}$, i.e. $A\ket{\lambda_{i}}=\lambda_{i}\ket{\lambda_{i}}$, $i=1,2$, so $A^{-1}A\ket{\lambda_{i}}=\ket{\lambda_{i}}=\lambda_{i}A^{-1}\ket{\lambda_{i}}$, thus the two eigenvalues of $A^{-1}$ are $\lambda_{1}^{-1}$ and $\lambda_{2}^{-1}$ (for $\lambda_{i}\neq 0$). If $|\lambda_{1}|<1$ and $|\lambda_{2}|<1$ (or, $|\lambda_{1}^{-1}|<1$ and $|\lambda_{2}^{-1}|<1$), we will have a decaying solution for $\phi_{N,i}$ (or $\phi_{N,N+1-i}$), $i=1,2,\cdots,N$, in the thermodynamic limit $N\to \infty$. This corresponds to $|\mu|<2w$ and $\Delta\neq 0$. For general values of $w_{j}$, $\Delta_{j}$ and $\mu_{j}$, Ref.$^{36}$ proves that if $w_{j}$ and $\Delta_{j}$ are sign-ordered i.e. sign$(\Delta_{j}w_{j})$=sign$(\Delta_{j+1}w_{j+1})$, and $|\mu_{j}|<\max(|w_{j-1}|,|\Delta_{j-1}|)$, then the chain has zero-energy Majoarana fermions. The solution~(4) can be verified by substituting it into Eq.~(\ref{transfer-phi-psi}). Actually, there is another solution: $\mu_{1}=\mu_{N}=0,\,\,\Delta_{1}=-w_{1},\,\,\Delta_{N-1}=-w_{N-1}$, corresponding to the absence of $d_{2}$ and $d_{2N-2}$ in Eq.~(3). The two MFs form a zero-energy edge mode that is similar to Eq.~(5). Further analysis for this solution is also very similar to that for Eq.~(4) and omitted.

Next, we prove the parity equality: $\prod_{j=1}^{N}(1-2c_{j}^{\dagger}c_{j})=\prod_{j=1}^{N}(1-2\tilde{b}_{j}^{\dagger}\tilde{b}_{j})$. The Hamiltonian~(3) can be written as $H=\frac{i}{4}\sum_{l,m}A_{l,m}d_{l}d_{m}$. The coefficient matrix $A$ is block diagonalized by a $2N\times2N$ real orthogonal matrix $W$: $A_{j,k}=\sum_{m,n}W^{T}_{j,m}\Lambda_{m,n}W_{n,k}$, where $\Lambda$ is a block diagonal matrix: $\Lambda_{2k-1,2k}=-\Lambda_{2k,2k-1}=\lambda_{j}$ (other matrix elements are zero). Note that $WW^{T}=W^{T}W=I$ so that $(\det W)^{2}=\det W\det W^{T}=\det(WW^{T})=1$. When $\det W=1$, the parity equality holds$^{38}$. For the parameters in Eq.~(4), we have the zero-energy mode in Eq.~(5). The corresponding $W$ will be $W_{j,1}=\delta_{j,2N-1}$, $W_{j,2N}=W_{2N,j}=\eta\delta_{j,2N}$, $W_{2N-1,j}=\delta_{j,1}$ for $1\leq j\leq 2N$. While $W_{j,k}=(W_0)_{j,k-1}$ for $1\leq j\leq 2N-2$, $2\leq k\leq 2N-1$. Here $\eta=\det W_{0}$ is defined in Eq.~(5). Therefore, $\det W$ can be calculated, according to the definition of determinant, as $\det W=\eta\det W_{0}=\eta^{2}=1$. This concludes the proof.

\section*{Ground states in real space of the chain}\label{GS-real-space}
In real space of the chain, $\ket{G_{1}}$ can be written as
\ba\label{ground-state-real-space}
\ket{G_{1}}=\sum_{i,j=0}^{1}z_{ij}\ket{ij}_{1,N}\ket{\Psi_{ij}},
\ea
where we have grouped the states for the sites $1$ and $N$ together (assume $N\geq 3$ first), and $\ket{\Psi_{ij}}$ is the state for the sites from $2$ to $N-1$. We notice that because $\ket{G_{1}}$ has an even parity (i.e. it is a superposition of the states with an even number of electrons), the states $\ket{\Psi_{00}}$ and $\ket{\Psi_{11}}$ also have an even parity, while the states $\ket{\Psi_{01}}$ and $\ket{\Psi_{10}}$ have an odd parity.

The state $\ket{G_{1}}$ satisfies $\tilde{b}_{N}\ket{G_{1}}=0$. Using Eq.~(5), we have $\tilde{b}_{N}\ket{G_{1}}$ equal to

\ba
\ket{00}_{1,N}(z_{10}\ket{\Psi_{10}}-\eta z_{01}\ket{\Psi_{01}})+\ket{01}_{1,N}(z_{11}\ket{\Psi_{11}}-\eta z_{00}\ket{\Psi_{00}})\nonumber\\
+\ket{10}_{1,N}(z_{00}\ket{\Psi_{00}}-\eta z_{11}\ket{\Psi_{11}})+\ket{11}_{1,N}(z_{01}\ket{\Psi_{01}}-\eta z_{10}\ket{\Psi_{10}}).\nonumber
\ea
We can set $z_{10}=\eta z_{01}\equiv x_{1}/\sqrt{2}$, $z_{00}=\eta z_{11}\equiv x_{0}/\sqrt{2}$, $\ket{\Psi_{10}}=\ket{\Psi_{01}}\equiv \ket{S_{e}}$, $\ket{\Psi_{00}}=\ket{\Psi_{11}}\equiv \ket{S_{o}}$ (note that $\eta^{2}=1$). Therefore, $\ket{G_{1}}$ equals
\ba\label{appendix-real-space-state}
x_{0}\frac{\ket{00}_{1,N}+\eta \ket{11}_{1,N}}{\sqrt{2}}\ket{S_{e}}+x_{1}\frac{\ket{10}_{1,N}+\eta \ket{01}_{1,N}}{\sqrt{2}}\ket{S_{o}},
\ea
We note that $\bra{G_{1}}c_{N}^{\dagger}c_{1}^{\dagger}c_{1}c_{N}\ket{G_{1}}=\frac{1}{2}|x_{0}|^2$. For the homogeneous chain ($\Delta_{j}=\Delta$, $w_{j}=w$, $\mu_{j}=\mu$), if we further have $\Delta=w>0, \mu=0$, the Hamiltonian~(3) reduces to $H=iw\sum_{j=1}^{2N-1}d_{2j}d_{2j+1}$ and the new Dirac fermion operators that diagonalize the Hamiltonian~(3) are $\tilde{b}_{j}=(c_{2j}+ic_{2j+1})/2$, $\tilde{b}_{j}^{\dagger}=(c_{2j}-ic_{2j+1})/2$, for $j=1,2,\cdots,N-1$, ($\eta=1$, see Ref.$^{38}$). These relations together with Eq.~(2) and~(5) can be used to solve for $c_{1}$ and $c_{N}$. $c_{1}=\frac{1}{2}(\tilde{b}_{N}+\tilde{b}_{N}^{\dagger}+i\tilde{b}_{1}+i\tilde{b}_{1}^{\dagger})$,
$c_{N}=\frac{1}{2}(\tilde{b}_{N}-\tilde{b}_{N}^{\dagger}-i\tilde{b}_{N-1}+i\tilde{b}_{N-1}^{\dagger})$.
So $\bra{G_{1}}c_{N}^{\dagger}c_{1}^{\dagger}c_{1}c_{N}\ket{G_{1}}$ can be calculated by using these expressions and Eq.~(6), and the result is $\frac{1}{4}$. Thus, $x_{0}$ can be chosen to be $\frac{\sqrt{2}}{2}$, and  $x_{1}$ is determined by normalization condition of the state~(8): $|x_{0}|^2+|x_{1}|^2=1$, so $x_{1}=\frac{\sqrt{2}}{2}$. For general values of $\Delta_{j},w_{j},\mu_{j}$, we only have Eq.~(2) and~(5), but $\tilde{b}_{j}$ ($j=1,2,\cdots,N-1$) is unknown, which is a function of $c_{k},c_{k}^{\dagger}$ ($k=1,2,\cdots,N$) and can be obtained by diagonalizing the coefficient matrix in the Hamiltonian~(1). Then, we can solve for $c_{1}$ and $c_{N}$ to determine $x_{0}, x_{1}, \eta$ in the state~(8). The state $\ket{G_{2}}=\tilde{b}_{N}^{\dagger}\ket{G_{1}}$ by using Hermitian conjugate of Eq.~(5). For the situation $N=2$, the calculation is similar, where the two ground states are those before $\ket{S_{e}}$ and $\ket{S_{o}}$ in Eq.~(\ref{appendix-real-space-state}).

\section*{Estimation of $J$}\label{estimation-J}
The coupling between the microwave and the spin of the quantum dots in our proposal is analogous to the electron spin resonance which is usually weak as compared with the coupling through the electric dipole moment and thus ignored. However, in our experiment, this coupling is significant due to the presence of strong spin-orbit coupling in the quantum dots. We shall estimate the coupling strength $J$ below. Note that the microwave is applied to the two boundary quantum dots which are decoupled from the inner part of the chain. Therefore, the superconducting proximity effect are not present in the measurement process.

The eigenstates and energy eigenvalues for the two spin-split levels of a single quantum dot are$^{36}$
\ba\label{spin-split-levels}
\ket{\psi_{\pm}}&=&e^{-i(\frac{\pi x}{l_{so}}+\eta)\sigma_{y}}\ket{\psi_{0},\sigma_{z}=\pm 1},\nonumber\\
E_{\pm}&=&E_{0}-\frac{\pi^{2}}{2m^{*}l_{so}^{2}}\pm V_{z}\sqrt{c_{0}^{2}+s_{0}^{2}},
\ea
where $l_{so}=\frac{\pi}{m^{*}\alpha}$ ($m^{*}$ the electron's effective mass, $\alpha$ the Rashba spin-orbit coupling strength), $\tan\eta=c_{0}/s_{0}-\sqrt{1+(c_{0}/s_{0})^{2}}$ with $c_{0}=\bra{\psi_{0}}\cos\frac{2\pi x}{l_{so}}\ket{\psi_{0}}$, $s_{0}=\bra{\psi_{0}}\sin\frac{2\pi x}{l_{so}}\ket{\psi_{0}}$, $E_{0}$ and $\ket{\psi_{0}}$ are the eigen-energy and the spatial part of the eigen wave function without considering the magnetic field and the spin-orbit coupling, and $V_{z}$ is the Zeeman potential.

With the measurement setup in Fig.~5, the Hamiltonian for a single quantum dot and the microwave changes to $\hbar\omega_{m}a^{\dagger}a+\frac{1}{2m^{*}}(\mathbf{p}+e\mathbf{A})^{2}-e\phi_{m}+\alpha\hat{\mathbf{z}}\cdot[(\mathbf{p}+e\mathbf{A})\times\boldsymbol{\sigma}]+V(\mathbf{r})$, where $a^{\dagger}$($a$) is the creation (annihilation) operator of the microwave photons with frequency $\omega_{m}$, $\mathbf{p}$ is the momentum of the electron, $\mathbf{A}$ and $\phi_{m}$ are the vector and scalar potentials of the microwave, $\boldsymbol{\sigma}=(\sigma_{x},\sigma_{y},\sigma_{z})$ is a vector of Pauli matrices, $V(\mathbf{r})$ is the confining potential. The term $-e\phi_{m}\sim e\mathbf{r}\cdot(\mathbf{E}+\partial_{t}\mathbf{A})$ describes the interaction of the electromagnetic filed with the dipole moment of the electron. This term is considered for coupled double quantum dots\cite{TLR-QD-1}, but it can be ignored here for a single quantum dot (or two decoupled quantum dots) because the only possible transition is that between the two energy levels in Eq.~(\ref{spin-split-levels}) which is electric-dipole forbidden ($\bra{\psi_{+}}\mathbf{r}\ket{\psi_{-}}=0$). An exception is to use a third energy level to induce Raman transition\cite{Raman-transition}, but this method will not be considered here due to its incapability of realizing the dispersive coupling. One could also consider this third level and $\ket{\psi_{-}}$ forming a two-level system which allows transition through the electric dipole (usually with optical frequency). This method is feasible for achieving the dispersive coupling, but transitions involving other levels (e.g. from the aforementioned third level to $\ket{\psi_{+}}$) must be suppressed, otherwise spin-photon entanglement will be generated\cite{spin-photon-etg}. Instead, when restricted to the subspace spanned by $\ket{\psi_{\pm}}$, the relevant part describing the electron-microwave interaction in the Hamiltonian is $H_{r}=\frac{e}{m^{*}}(p_{x}A_{x}+p_{y}A_{y}+p_{z}A_{z})+e\alpha(\sigma_{y}A_{x}-\sigma_{x}A_{y})$. We assume that $\bra{\psi_{0}}\mathbf{p}\ket{\psi_{0}}=0$, and notice that $\bra{\psi_{+}}p_{x}\ket{\psi_{-}}=i m^{*}\alpha$. 
The electromagnetic gauge is chosen to be\cite{spin-microwave,TLR-QD-1} $\mathbf{A}=\frac{\mathbf{k}\times\mathbf{B_{0}}}{k^{2}}(a+a^{\dagger})$, $\phi_{m}=\tilde{\phi}_{m}(a+a^{\dagger})$, $\tilde{\phi}_{m}=\frac{l}{l_{0}}\sqrt{\frac{\hbar\omega_{m}}{c_{tot}}}$, where $\mathbf{k}$ is the wave vector of the microwave, $\mathbf{B_{0}}$ is the rms vacuum fluctuations of the magnetic field of the microwave, $c_{tot}$ is the total capacitance of the transmission line resonator (TLR), $a$ ($a^{\dagger}$) is the annihilation (creation) operator of the microwave photons, $l$ is the coordinate along the electric field lines, and $l_{0}$ is the distance between the two planes facing the quantum dots (see Fig.~7). The second quantization of $H_{r}$ with the rotating-wave approximation is $(f_{j}^{\dagger}c_{j}\bra{\psi_{+}}H_{r}\ket{\psi_{-}}+h.c.)=Jf_{j}^{\dagger}c_{j}a+h.c.$.
\ba\label{J-derivation}
J=e\alpha c_{\eta}|\mathbf{B_{0}}|/k,
\ea
where $c_{\eta}=-\bra{\psi_{0}}\cos(\frac{2\pi x}{l_{so}}+2\eta)\ket{\psi_{0}}$. Experimentally, $\alpha=\pi/(m^{*}l_{so})$ with $m^{*}=0.023 m_{e}$ for the InAs quantum dot, $l_{so}\sim100$ nm, and $|\mathbf{B_{0}}|\sim|\nabla_{l}\tilde{\phi}_{m}|/c=\frac{1}{l_{0}c}\sqrt{\frac{\hbar\omega_{m}}{c_{tot}}}$ ($c$ the speed of light), $\omega_{m}/2\pi\sim 10^{11}$ Hz, $c_{tot}\sim 1$ pF, $l_{0}\sim5$ $\mu$m, $k=2\pi/\lambda\sim 2\times10^{3}$ m$^{-1}$ and assume $\braket{x}{\psi_{0}}=\sqrt{\frac{2}{L}}\sin(\frac{\pi x}{L})$ with $L\sim120$ nm the width of the quantum dot in $x$ direction, $J/2\pi$ can be tuned to $~200$ MHz. The phase $\phi$ in Eq.~(11) can be generated through the time evolution $e^{it\delta f_{j}^{\dagger}f_{j}}f_{j}^{\dagger}e^{-it\delta f_{j}^{\dagger}f_{j}}=f_{j}^{\dagger}e^{it\delta}$ with $t=(\phi+2n\pi)/\delta$, $n$ an integer. If we ignore the spin-orbit coupling, but instead consider the direct interaction of the microwave's magnetic field with the electron's spin via the Zeeman effect: $H_{z}=-\boldsymbol{\mu}\cdot\mathbf{B}=\frac{e}{2m^{*}}\boldsymbol{\sigma}\cdot\mathbf{B_{0}}(a+a^{\dagger})$. The corresponding coupling strength after the second quantization and the rotating-wave approximation are performed will be $J_{z}=e|\mathbf{B_{0}}|/(2m^{*})$. The ratio $J/J_{z}=2\alpha c_{\eta}m^{*}/k=c_{\eta}\lambda/l_{so}\sim 10^{4}$, indicating that the spin-microwave interaction will be considerably enhanced when the spin-orbit coupling is significant. We notice that this enhancement has already been pointed out in Ref.$^{47}$ of the main text, where the spin-photon coupling is realized in the interacting double quantum dots via the spin-orbit interaction. The coupling strength there $\sim0.4$ MHz (see F.2. in the Supplemental Material of Ref.$^{47}$) is much smaller than our value $\sim200$ MHz. The main reason is that the spin rotation of the wave function in Ref.$^{47}$ is along the circumference of the carbon nanotube. For this configuration, the overlap between the microwave's vector potential and the electron's momentum vector with a varying direction (tangential to the circumference) is rather limited.

\begin{figure}
\setcounter{figure}{6}
\centering
\includegraphics[width=0.35\textwidth,height=0.26\textwidth]{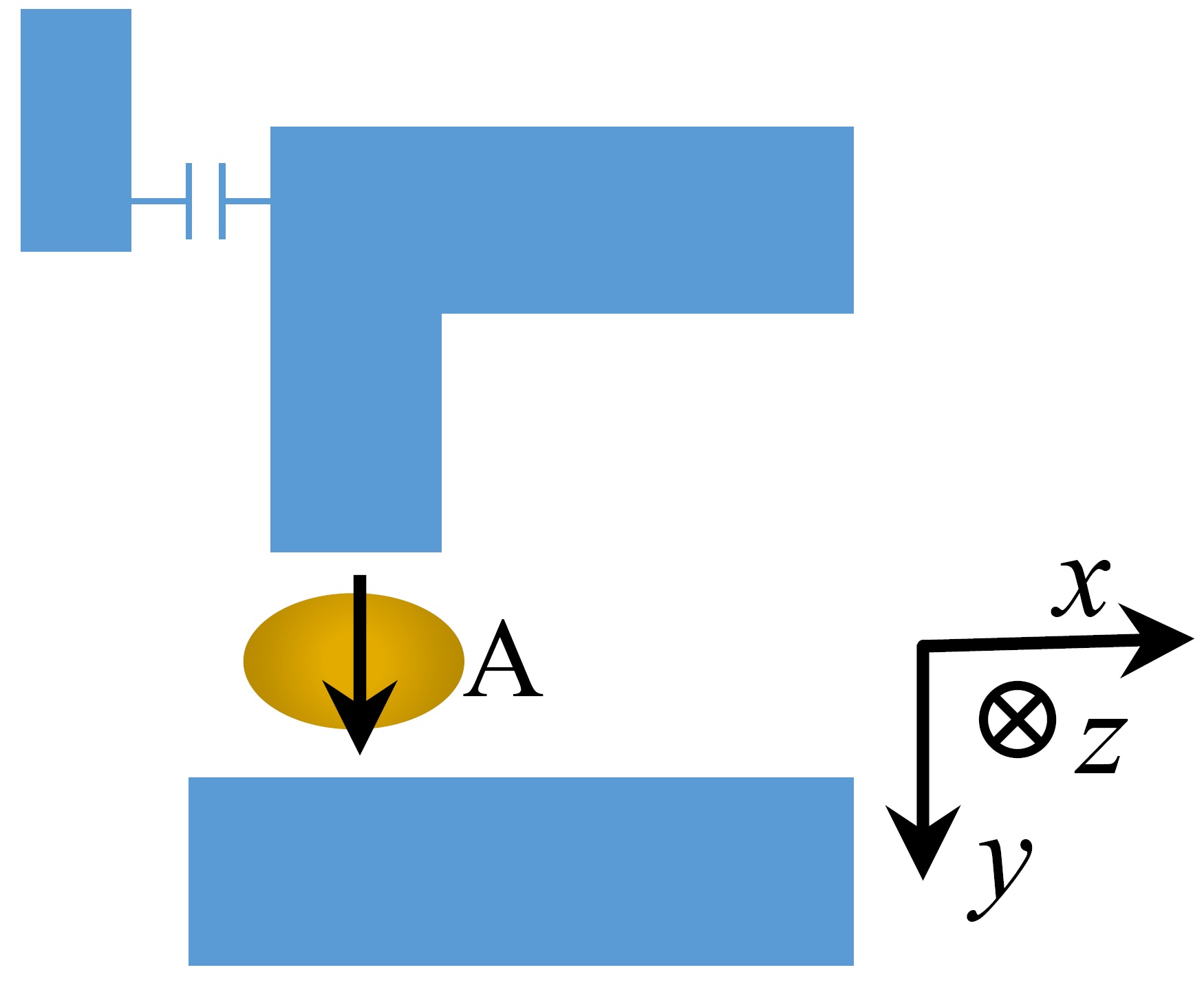}
\caption{\textbf{The detailed diagram of the measuring setup.}} \label{Fig-7-measure-detail}
\end{figure}

\section*{Teleportation}\label{teleport-sec}
The state~(9) will be useful for performing quantum teleportation if its parity is determined through the microwave measurement. Suppose the boundary sites are in the state: $\frac{1}{\sqrt{2}}(\ket{00}_{1,N}+\ket{11}_{1,N})$. We have an extra fermionic site with the index $0$. The quantum state of this site may be correlated with its environment. Denote the state of the site $0$ and its environment as $\ket{\chi_{0}}=y_{0}\ket{0}_{0}\ket{\phi_{0}}+y_{1}\ket{1}_{0}\ket{\phi_{1}}$, where $\ket{0}_{0}$ ($\ket{1}_{0}$) means that there is no (one) fermion in the site $0$, $y_{0}$ and $y_{1}$ are the amplitudes, and $\ket{\phi_{0}}$ and $\ket{\phi_{1}}$ are some states of the environment. The task is to teleport the state in the site $0$ to the site $N$. As we know, one need to perform Bell measurement. This seems not possible, but we can realize it indirectly through basis transformation in the Hilbert space of the site $0$ and $1$.

The initial state is
\ba
\ket{\phi}=\ket{\chi_{0}}\frac{\ket{00}_{1,N}+\ket{11}_{1,N}}{\sqrt{2}}
\ea
Then the interaction (hopping and pairing) between the sites $0$ and $1$ is switched on, and their chemical potential is finely tuned to be zero. The Hamiltonian is
\ba\label{bell-Hamiltonian}
H_{01}=-w_{0}(c_{0}^{\dagger}c_{1}+c_{1}^{\dagger}c_{0})+\Delta_{0}(c_{0}c_{1}+c_{1}^{\dagger}c_{0}^{\dagger})
\ea
Choose the hopping amplitude $w_{0}$ and the superconducting gap $\Delta_{0}$ to satisfy $w_{0}=\Delta_{0}>0$. The time evolution operator is $\exp(-itH_{01})$. Setting $t=t_{0}\equiv\frac{\pi}{4\Delta_{0}}$, we have
\ba\label{teleport-state}
e^{-it_{0}H_{01}}\ket{\phi}=\displaystyle\frac{1}{2}[\ket{00}_{01}(y_{0}\ket{0}_{N}\ket{\phi_{0}}+iy_{1}\ket{1}_{N}\ket{\phi_{1}})\hspace{0.45cm}\nonumber\\
+\ket{01}_{01}(y_{0}\ket{1}_{N}\ket{\phi_{0}}+iy_{1}\ket{0}_{N}\ket{\phi_{1}})\hspace{0.5cm}\nonumber\\
+\ket{10}_{01}(iy_{0}\ket{1}_{N}\ket{\phi_{0}}+y_{1}\ket{0}_{N}\ket{\phi_{1}})\hspace{0.5cm}\nonumber\\
+\ket{11}_{01}(iy_{0}\ket{0}_{N}\ket{\phi_{0}}+y_{1}\ket{1}_{N}\ket{\phi_{1}})],\hspace{0.25cm}
\ea
where we have written the states of the sites $0$ and $1$ first, and then the states of the site $N$ and the environment of the site $0$. It can be seen that we have four results when measuring the site $0$ and $1$ in the number basis: $\ket{jk},\,j,k\in\{0,1\}$ (see Ref.$^{28}$ for the experimental realization of charge measurement). For each result, the corresponding state for the site $N$ and the environment of the site $0$ is equivalent to the original state $\ket{\chi_{0}}$ up to a local unitary transformation (gate) on the site $N$. The results $00$ and $11$ involve phase gates which are fulfilled by applying electric voltage, while the results $01$ and $10$ involve bit-flip gates which are realized by coupling the site $N$ to a chain supporting zero-energy edge mode $\tilde{b}_{m}^{\dagger}\tilde{b}_{m}$. The chain has a Hamiltonian similar to Eq.~(1) with parameters in Eq.~(4), but $N$ there is all replaced by another length $m$. The inner sites of a sub-chain in Fig.~1 can be chosen to serve as this chain. The coupling between the site $N$ and the chain is$^{46}$: $H_{f}=\kappa(c_{N}^{\dagger}-c_{N})(\tilde{b}_{m}^{\dagger}+\tilde{b}_{m})$, where $\kappa$ denotes the coupling strength. We have $\exp(-itH_{f})\ket{0}_{N}\ket{G_{1}}=-i\ket{1}_{N}\ket{G_{2}}$ and $\exp(-itH_{f})\ket{1}_{N}\ket{G_{1}}=-i\ket{0}_{N}\ket{G_{2}}$ when $t=\pi/(2\kappa)$, realizing the bit-flip gate. Here $\ket{G_{1}}$ and $\ket{G_{2}}$ are the ground states of the chain (see Eq. (6) with $N$ replaced by $m$), and their positions can be interchanged to obtain the other two Eqs. of time evolution.

For the other maximally entangled state $(\ket{10}_{1,N}+\ket{01}_{1,N})/\sqrt{2}$, the discussion is very similar. We only need to flip the state of the site $N$ in Eq.~(\ref{teleport-state}) to obtain the result. To suppress the charge noise, it is preferred to use the basis of $\ket{10}$ and $\ket{01}$ rather than $\ket{00}$ and $\ket{11}$ of the double quantum dots for encoding the qubit. In this situation, the state $(\ket{00}_{1,N}+\ket{11}_{1,N})/\sqrt{2}$ and the pairing interaction in (\ref{bell-Hamiltonian}) should be avoided. When $\Delta_{0}$ is set to be zero in (\ref{bell-Hamiltonian}), it can be verified that the teleportation succeeds with the probability $1/2$. 

\end{document}